\documentclass[aps,prd,12pt,citeautoscript,superscriptaddress,showpacs,scrartcl]{revtex4-1}
\usepackage[utf8x]{inputenc}

\usepackage{xcolor}
\usepackage[colorlinks=true, pdfstartview=FitV,linkcolor=blue, citecolor=blue, urlcolor=blue]{hyperref}

\usepackage[bottom]{footmisc}

\usepackage{graphicx}

\makeatletter
\newcommand*\bigcdot{\mathpalette\bigcdot@{.5}}
\newcommand*\bigcdot@[2]{\mathbin{\vcenter{\hbox{\scalebox{#2}{$\m@th#1\bullet$}}}}}
\makeatother

\newcommand{\be}{\begin{eqnarray}}
\newcommand{\ee}{\end{eqnarray}}
\newcommand{\bea}{\begin{eqnarray}}
\newcommand{\eea}{\end{eqnarray}}

\newcommand{\nn}{\nonumber}
\newcommand{\ssigma}{\mu}

\usepackage{natbib}
\usepackage{graphicx}
\usepackage{amsmath}
\usepackage{amssymb}
\usepackage{mathtools}
\usepackage{tensor}
\usepackage{accents}
\usepackage{cancel}
\usepackage{titlesec}
\usepackage{subfigure}
\usepackage[a4paper]{geometry}
\usepackage{psfrag}
\usepackage{latexsym,array,theorem,mathrsfs,subfigure,bm,float}
\usepackage{amsfonts}
\usepackage{amssymb}

\newcommand{\kk}{{\bf k}}

\newcommand{\OOmega}{{\bf U}}

\newcommand{\W}{{\bf w}}

\titleformat{\paragraph}
{\normalfont\normalsize\bfseries}{\theparagraph}{1em}{}
\titlespacing*{\paragraph}
{0pt}{3.25ex plus 1ex minus .2ex}{1.5ex plus .2ex}

\begin{document}

\title{Stationary generalizations for the Bronnikov-Ellis wormhole and for  the vacuum ring wormhole}



\author{Mikhail~S.~Volkov}
\email{volkov@lmpt.univ-tours.fr}
\affiliation{
Institut Denis Poisson, UMR - CNRS 7013, \\ Universit\'{e} de Tours, Parc de Grandmont, 37200 Tours, France}

\affiliation{
Department of General Relativity and Gravitation, Institute of Physics,\\
Kazan Federal University, Kremlevskaya street 18, 420008 Kazan, Russia
}

\begin{abstract} 

\vspace{1 mm}

We analyze possibilities  to obtain a globally regular 
stationary generalization 
for  the ultrastatic wormhole with a repulsive scalar field found by Bronnikov and by Ellis in 1973.  
The extreme simplicity of this static solution  
suggests that its spinning version   could be obtainable  analytically and should be globally regular, but 
no such generalization  has been 
found.  We  analyze the problem and find that the difficulty 
originates in the vacuum theory, 
since the scalar field can be eliminated within the Eris-Gurses procedure. 
 The problem 
then reduces  to constructing the spinning generalization for the 
vacuum wormhole sourced by a thin  ring of  negative tension. 
Solving the vacuum Ernst equations determines  the $g_{00}$, $g_{0\varphi}$ 
metric components and hence   the AMD mass ${\rm M}$ and angular momentum ${\rm J}$,
all of these being specified by  the ring source.  The scalar field can be  
included  into consideration afterwords, but this only affects 
$g_{rr}$ and $g_{\vartheta\vartheta}$ without changing  the rest. 
Within this approach, we analyze a number of 
exact stationary generalizations for the  wormhole, but none of them are satisfactory. 
However, the perturbative expansion around  the static vacuum background  contains only bounded functions
and presumably converges to an 
exact solution.  Including  the scalar field screens 
the singularity at the ring source and renders the geometry regular.  This solution 
describes a globally regular spinning wormhole with two asymptotically flat regions. Even though the source 
itself is screened and not visible, 
the  memory of it remains in $g_{00}$ and $g_{0\varphi}$ and accounts for 
the ${\rm M}\propto {\rm J}^2$ relation typical for a rotating extended source. 
Describing stationary spacetimes  with an extended 
source is a complicated problem, which presumably explains the difficulty in finding the solution.

\end{abstract} 

\maketitle

\section{Introduction}

Wormholes are bridges or tunnels  between
different universes or different parts of the same universe. They were first introduced 
by Einstein and Rosen  (ER) \cite{Einstein:1935tc}, who noticed that the Schwarzschild black hole 
actually has two exterior  regions connected by a spacelike bridge. 
One can also discuss 
traversable wormholes accessible for ordinary classical particles 
or light  (see \cite{Visser:1995cc} for a review), but 
their existence requires  \cite{Friedman:1993ty,Hochberg:1998ii} that the Null Energy Condition (NEC) 
must be violated. 
Therefore,  traversable wormholes are possible if only the 
energy density becomes negative, for example due to vacuum polarization 
 \cite{Morris:1988tu} or due to   exotic matter 
\cite{Bronnikov:1973fh,Ellis:1973yv}.

Since the energy is normally supposed to be positive, the traversable wormholes 
were for a longtime considered as something  odd. 
The situation changed after the discovery of the cosmic acceleration
\cite{1538-3881-116-3-1009,0004-637X-517-2-565}, which  invoked a large number of 
alternative gravity models in which the energy is not necessarily positive definite. 
Wormholes have been found 
in many such theories, as for example 
in the Gauss-Bonnet theory \cite{Kanti:2011jz,Cuyubamba:2018jdl}, 
in the brainworld  models 
\cite{Bronnikov:2002rn}, 
in  theories with 
non-minimally coupled fields 
\cite{Sushkov:2011jh},
in massive (bi)gravity \cite{Sushkov:2015fma}, etc. 
As a result, wormholes  have become  
 quite popular nowadays. 

We do not intend  to argue that wormholes actually exist, neither shall we advocate the opposite viewpoint. 
We are merely interested in the problem of constructing solutions describing 
{\it spinning} wormholes in the theory with  a gravity-coupled 
phantom scalar field $\Phi$. This theory presents 
 very simple and certainly best known wormhole solutions 
found in 1973 
by Bronnikov and by Ellis (BE) 
 \cite{Bronnikov:1973fh,Ellis:1973yv}. 
Their simplest version is 
\be               \label{BE0}
d{\rm s}^2=-d{\rm t}^2+d{\rm r}^2+({\rm r}^2+\mu^2)(d\vartheta^2+\sin^2\vartheta d\varphi^2)
\ee
with the scalar field $\Phi=\arctan({\rm r}/\mu)$. The parameter $\mu$ determines the size 
of the wormhole throat, 
the radial coordinate ${\rm r}\in(-\infty,\infty)$, and the limits ${\rm r}\to\pm \infty$ 
correspond to two asymptotically flat regions connected through the wormhole throat.

The theory also admits exact axially symmetric solutions describing  superpositions of several wormholes 
\cite{Clement:1983ic,Clement:2015lul,Egorov:2016rfr,Gibbons:2017jzk}, as well as 
solutions describing 
axially symmetric deformations  of a single wormhole  
 \cite{Gibbons:2016bok,Gibbons:2017jzk}. The latter are all singular, and one can 
prove that the BE solutions do not admit 
globally regular generalizations in the {\it static} sector \cite{Yazadjiev:2017twg}.


At the same time, nothing forbids the existence of 
globally regular {\it stationary} generalizations for the BE solutions which 
would describe spinning wormholes. 
The extreme simplicity of the  solution \eqref{BE0}  suggests that its stationary 
version could  be easily obtainable analytically, and  it is natural to expect this spinning solution 
to be globally regular. 
However,  even now, almost 50 years later, 
this solution is still unknown. 

Spinning wormholes are in fact often discussed in the literature (see, e.g. \cite{Deligianni:2021ecz}),  but   what 
is usually   meant are not exact solutions  but some model geometries 
\cite{Teo:1998dp}. Exact stationary solutions  are also  known, 
but they show singularities, for example of the NUT type  \cite{Clement:1983ib} or present  other 
problems 
\cite{Matos:2009au}. At the same time, 
there exist  perturbative (up to the second order  terms) \cite{Kashargin:2007mm,Kashargin:2008pk} and numerical 
\cite{Kleihaus:2014dla,Chew:2016epf} indications in favour of existence of 
{\it globally regular} spinning generalizations for the BE wormholes. However, 
their analytical form is unknown.

Exact stationary solutions may sometimes be obtained by applying the   generating methods
like dualities \cite{Clement:1997tx,Clement:1998nk},  \cite{Bogush:2020lkp} or by using 
some other tricks \cite{Newman:1965tw}, but 
none of these methods help to construct regular spinning wormholes. 
Therefore, 
in what follows we are trying to anlalyze the situation to understand why the problem 
is so difficult and what can be done.  It seems that the problem originates already in the vacuum theory 
and can be summarized as follows. 

Already before the BE discovery, it was known that the vacuum General Relativity  admits the 
{\it ring wormholes} described by the {\it oblate} metrics of Zipoy and Vorhees \cite{Zipoy,Voorhees:1971wh},
whose simplest version is 
\be               \label{BE1}
d{\rm s}^2=-d{\rm t}^2+\frac{{\rm r}^2+\mu^2\cos^2\vartheta}{{\rm r^2}+\mu^2}\left[d{\rm r}^2
+({\rm r}^2+\mu^2) d\vartheta^2\right] +({\rm r^2}+\mu^2)\sin^2\vartheta d\varphi^2
\ee
with ${\rm r}\in(-\infty,\infty)$. Although looks complicated, this would be just the Minkowski  metric expressed 
in spheroidal coordinates, 
if the radial coordinate was restricted to ${\rm r}\in[0,\infty)$. 
For
${\rm r}\in(-\infty,\infty)$ 
this metric is only locally flat and  describes a wormhole made of two copies of 
Minkowski space glued to each other through the disk bounded by the circle ${\rm r=0}$, $\vartheta=\pi/2$ 
in the equatorial plane. The circle carries a distributional singularity of the Ricci tensor that 
can be viewed as a singular matters source: a ring (loop) made of a cosmic string of {\it negative} tension
with the negative angle deficit $-2\pi$ \cite{Gibbons:2016bok,Gibbons:2017jzk}. 
As was noticed in \cite{Gibbons:2017djb}, the metric 
\eqref{BE1} is a special limit of the Kerr geometry.   All of this will be explained below. 

Now, the BE solution \eqref{BE0} can be  obtained from the ring metric \eqref{BE1}. 
Assuming that $\Phi=\Phi({\rm r})$, 
the scalar field equation $\Delta\Phi=0$ has the same form in both metrics and so has the same solution.
Adding $\Phi$ as the  source  to the Einstein equations only modifies (removes) the conformal factor in front 
of the ${\rm r},\vartheta$ part of the metric \eqref{BE1}, after which the metric  reduces to \eqref{BE0}. 
We shall call  this  procedure ``dressing", hence 
the BE wormhole is the ring wormhole  ``with scalar  dressing".  A similar procedure works also in the stationary 
case \cite{Eris}, and  finding a stationary generalization 
for the BE solution  \eqref{BE0}  reduces to solving the same  problem for the vacuum 
ring metric \eqref{BE1}.  The vacuum Ernst equations determine the $g_{00}$, 
$g_{0\varphi}$, $g_{\varphi\varphi}$  metric components  yielding the ADM mass 
and angular momentum, all of  these being  insensitive 
to the scalar field. 
The latter only modifies 
$g_{rr}$, $g_{\vartheta\vartheta}$ not affecting the rest. As a result, 
all the essential features of the system are encoded already in the vacuum theory. 

Therefore, all we need to do is to solve the vacuum Ernst  equations. 
This is not always easy but  still simpler than to directly attack 
the full system of coupled Einstein and scalar field equations as was done in 
\cite{Kashargin:2007mm,Kashargin:2008pk} and in 
\cite{Kleihaus:2014dla,Chew:2016epf}.  
Following this logic, we apply in what follows  the procedure based on 
vacuum Ernst equations  to construct 
stationary solutions with the  scalar field. 
Our ultimate goal  is to try and  possibly obtain the globally regular rotating wormholes exactly. 
We indeed obtain new 
exact solutions, but  these are not globally regular. 
At the same time, we construct the  perturbative expansion for the globally regular solution 
of the Ernst equations 
which determine the asymptotically flat stationary generalizations for both the ring 
wormhole \eqref{BE1} and for the BE wormhole \eqref{BE0}. However, promoting this perturbative solution 
to an exact one does not seem to be obvious. 

Intuitively, the explanation of the difficulty is that the ring metric \eqref{BE1} has en extended source -- 
the cosmic string loop. This source is hidden also in the spinning version of the regular BE solution \eqref{BE0}, 
although not directly visible there, being screened by the scalar. 
However,  since its $g_{00}$ and $g_{0\varphi}$ metric components are the same as for the ring wormhole, 
the  BE wormhole 
shows the same relation between the mass and angular momentum which is 
typical for a rotating extended source -- the ring. The spinning BE wormhole 
``knows" about this  source. However, 
finding a stationary 
solution with an extended source is  a more difficult problem than finding it for a pointlike source, say 
(the Kerr metric).  Therefore, although the spinning version of the BE wormhole 
can be constructed perturbatively or numerically, it may be not expressible in a compact analytical 
form.  This presumably explains why this solution has never been obtained, 
despite the apparent  simplicity of its static 
limit described by \eqref{BE0}. 

In what follows we describe  the approach based on solving the vacuum Ernst equations first 
and including  the scalar field afterwords.  We  consider a number of stationary 
generalizations for  the wormhole. The most obvious one is the Kerr 
metric, however, adding to it the scalar dressing yields a singular result. We then analyze 
a special ansatz reducing  the Ernst equations to a  harmonic equation. 
This yields  exact solutions which are ``almost" perfect but unfortunately are not globally regular. 
We then consider the perturbative expansion 
around the static ring metric \eqref{BE1} and find that it contains unbounded 
functions and hence is ill-defined. However, reformulating the Ernst equations in terms 
of the axial Killing vector instead of the timelike one yields a better result, 
and we explicitly construct the globally regular perturbative expansion 
up to the fourth order terms.  This expansion presumably converges to an 
exact solution describing the correct stationary generalization for the ring wormhole and for the 
BE wormhole. We finish by discussing 
 chances   to get this solution exactly.

\section{The theory}
\setcounter{equation}{0}

We consider the theory with a minimally coupled to gravity scalar field with a ``wrong" sign in front 
of the kinetic term. It is convenient to introduce from the very beginning the length scale ${\mu}$ and represent the 
line element as 
\be              \label{mu}
d{\rm s}^2={\mu}^2 g_{\mu\nu} dx^\mu dx^\nu\equiv \mu^2 ds^2,
\ee
where the metric $g_{\mu\nu}$ and coordinates $x^\mu$ are dimensionless; their dimensionful analogues will be 
denoted by roman symbols. 
The action of the theory  is 
\be              \label{1}
S=\frac12\, \mu^2 {\rm M}_{\rm Pl}^2\int \left( {R}+2\partial_\mu\Phi\partial^\mu\Phi  \right) \sqrt{-{g}} \,d^4{x} 
\equiv  \frac12\, \mu^2 {\rm M}_{\rm Pl}^2 \int {\cal L}_4 \sqrt{-{g}}\, d^4{x}\,,
\ee
 which yields upon varying the equations 
\be
R_{\mu\nu}=-2\partial_\mu\Phi\partial_\nu\Phi,~~~~~~\nabla_\mu \nabla^\mu\Phi=0. 
\ee
Assuming the system to be stationary, the metric is chosen in the Papapetrou form, 
\be                \label{2}
ds^2 =-e^{2U} \left(dt-w_k dx^k\right)^2+e^{-2U}\, h_{ik}\, dx^i dx^k\,,
\ee
where  the  Newtonian potential $U$, the rotation field $w_k$, and the 3-metric $h_{ik}$ 
depend on the spatial coordinates $x^k$. 
Inserting this to \eqref{1} yields
\be      
{\cal L}_4\sqrt{-g}  =      
{\cal L}_3\sqrt{h}+\text{total derivative}
\ee
with 
\be          \label{lag}
{\cal L}_3=\overset{(3)}{R}(h)-2(\partial U)^2+\frac14\, e^{4U}F_{ik}F^{ik}+2(\partial \Phi)^2, 
\ee
were  $F_{ik}=\partial_i w_k-\partial _k w_i$ and 
$(\partial U)^2=\partial_k U\partial^k U$ with the indices 
moved by $h_{ik}$. Varying this Lagrangian yields the equations
\begin{subequations}                     \label{eqs}
\begin{align}
\nabla^k\nabla_k U+\frac14\, e^{4U}\,F_{ik} F^{ik}=0\,,\label{F1} \\
{\nabla}_i\left(e^{4U} F^{ik}\right)=0\,,\label{F2}  \\
\nabla^k\nabla_k\Phi=0\,, \label{F3} \\
\overset{(3)}{R}_{ik}-2\,\partial_i U\partial_k U+\frac12\, e^{4U} F_{is}F_k^{~s}+2\,\partial_i \Phi\partial_k \Phi=0\,, \label{F4} 
\end{align}
\end{subequations}
where $\nabla_k$ is the covariant derivative with respect to $h_{ik}$. 

\subsection{Static case}

In the static case, when $F_{ik}=0$, the Lagrangian \eqref{lag} and equations \eqref{eqs} are invariant 
under global rotations, 
\be                \label{rot}
U\to U\,\cosh\alpha+\Phi\sinh\alpha,~~~~~\Phi\to \Phi\,\cosh\alpha+ U\sinh\alpha,~~~~~h_{ik}\to h_{ik}\,.
\ee
This allows one to generate non-trivial solutions from a vacuum seed metric. Let us see how this works.
The Schwarzschild metric of unit mass  can be described by 
\be             \label{Sch}
e^{2U}=\frac{x-1}{x+1},~~~~~~h_{ik}\, dx^i dx^k=dx^2+(x^2-1)\,( d\vartheta^2+\sin^2\vartheta d\varphi^2   ),~~~~~\Phi=0. 
\ee
Applying to this the transformation \eqref{rot} with $\cosh\alpha=\delta$ yields the solution with a non-trivial scalar,
\be              \label{xx}
ds^2&=&-\left(\frac{x-1}{x+1}\right)^{\delta} dt^2+\left(\frac{x+1}{x-1}\right)^{\delta} \left[dx^2+(x^2-1)\,( d\vartheta^2+\sin^2\vartheta d\varphi^2   )\right],\nn \\
\Phi&=&\frac12\sqrt{1-\delta ^2}\,\ln\left(\frac{x-1}{x+1}\right), 
\ee
which reduces back to \eqref{Sch}  if $\delta =1$. 
Performing the analytic continuation, 
\be              \label{xxx}
t\to it,~~~~~x\to ix,~~~~~\delta\to i\delta,~~
\ee
and also replacing $\mu\to i\mu$ in \eqref{mu}, 
the line element  $ds^2=d{\rm s}^2/\mu^2$ and the scalar become 
\be                   \label{BE}
ds^2=-e^{2\delta\, \Psi}dt^2+e^{-2\delta\,\Psi} [dx^2+(x^2+1)( d\vartheta^2+\sin^2\vartheta d\varphi^2   )],~
~~~~\Phi=\sqrt{1+\delta^2}\,\Psi, 
\ee
with $\Psi=\arctan(x)$. Setting $t={\rm t}/\mu$ and $x={\rm r}/\mu$, this describes 
the static BE wormholes   \cite{Bronnikov:1973fh,Ellis:1973yv},
whose ultrastatic version  \eqref{BE0} is obtained when $\delta\to 0$. 

We are looking for the  stationary generalization of these solutions. Unfortunately, the global symmetry \eqref{rot} 
is lost in the stationary case and   one cannot play the same game again 
and generate solutions with a non-trivial scalar field 
starting from the Kerr metric, say.

\subsection{Stationary  case}
In this case there exist other global symmetries. 
Defining the twist potential $\chi$ via 
\be             \label{twist}
\partial^i\chi=\frac{e^{4U}}{\sqrt{h}}\,  \epsilon^{ijk}\partial_j w_k\,,
\ee
whose integrability is insured by \eqref{F2}, 
the Lagrangian assumes the form 
\be
{\cal L}_3&=&\overset{(3)}{R}(h)-2(\partial U)^2-\frac12\, e^{-4U}(\partial \chi)^2+2(\partial \Phi)^2 \nn \\
&\equiv& \overset{(3)}{R}(h)-2\,{\cal G}_{AB}\,\partial_i Y^A\partial^i Y^B\,,
\ee
where the target space coordinates are $Y^A=(U,\chi,\Phi)$ and 
the target space metric is 
\be             \label{tar}
{\cal G}_{AB}\,d Y^A d Y^B=(\partial U)^2+\frac14\, e^{-4U}(\partial \chi)^2-(\partial \Phi)^2 \,.
\ee
Introducing the complex Ernst potential 
\be                \label{E}
{\cal E}=e^{2U}+i\chi\equiv \pm \frac{1-\xi}{1+\xi}\,
\ee
(we shall be choosing either plus or minus sign in this formula, depending on the context, because ${\cal E}\to-{\cal E}$ is a symmetry), 
one has 
\be
dU^2+\frac14\,e^{-4U} d\chi^2=\,\frac{d{\cal E} \bar{d{\cal E}}}{{\cal E}+\bar{{\cal E}}}=\,\frac{d\xi \bar{d\xi}}{(\xi\bar{\xi}-1)^2}\,,
\ee
which is the metric on the hyperbolic space $H^2$ (Lobachevsky plane). 
Therefore, 
the target space \eqref{tar} is the pseudo-euclidean direct product $H^2\otimes R^1$
with the same geometry as the one induced on the hyperboloid 
\be
X_0^2-X_1^2-X_2^2=\frac14
\ee
in the 4-dimensional space with the metric 
\be
dS^2=-d\Phi^2-dX_0^2+dX_1^2+dX_2^2.
\ee
Isometries of this space are the shifts 
\be
\Phi\to \Phi+\Phi_0
\ee
and the $H^2$ isometries, which can be represented in the form
\be
{\cal E}\to {\cal E}+i\alpha,~~~\frac{1}{\cal E}\to \frac{1}{\cal E}+i\beta\,,~~~~\xi\to e^{i\gamma}\xi.
\ee
These isometries  can be used to produce  solutions with a NUT charge, which however 
does   not help to construct globally regular spinning wormholes.

\subsection{Stationary  and axially symmetric case}

Let us choose the spatial coordinates as $x^k=(\rho,z,\varphi)$ 
and assume that nothing depends on $\varphi$. The 3-metric can be represented  in the form 
\be           \label{h}
dl^2=h_{ik}\, dx^i dx^k=e^{2k}\left(d\rho^2+dz^2\right)+\rho^2 d\varphi^2,
\ee
while $w_k dx^k=w\,d\varphi$ where $w,k$, as well as $U,\Phi$ depend only on $\rho,z$. 
This form of the metric is possible only in the vacuum theory, otherwise 
one should replace $\rho^2$ in front of $d\varphi^2$ by a function of $\rho,z$
(an introduction into the theory of stationary gravitational fields can be found, e.g., in \cite{Heusler1996}). 
However, since the Ernst 
equations considered below  correspond to the vacuum theory, 
the choice \eqref{h}  of the 3-metric  is legitimate.

The function $k$ drops out from the
first three equations in \eqref{eqs}, since one has,  for example, 
\be              \label{U0}
\nabla^k\nabla_k  U=\frac{1}{\sqrt{h}}\,\partial_i\left(\sqrt{h}h^{ik}\partial_k U\right)
=e^{-2k}\left(\partial_{\rho\rho}U+\frac{1}{\rho}\partial_\rho U+\partial_{zz}U\right)\equiv e^{-2k}\Delta U,
\ee
where $\Delta$ 
is the standard flat space Laplace operator expressed in cylindrical coordinates. 
As a result, the first two equations in \eqref{eqs} decouple from the rest and comprise a closed system 
\be
\Delta U+\frac{e^{4U}}{2\rho^2}\left[ (\partial_\rho w)^2+(\partial_z w)^2 \right]&=&0,~~\nn \\
\rho\, \partial_\rho\left(\frac{e^{4U}}{\rho}\partial_\rho w\right)+\partial_z \left(e^{4U}\partial_z w\right)&=&0, 
\ee
which can be written compactly as 
\be                      \label{Uw}
\Delta U+\frac{e^{4U}}{2\rho^2}(\vec{\nabla} w)^2=0,~~~~~\vec{\nabla}\left( \frac{e^{4U}}{\rho^2}\vec{\nabla} w  \right)=0. 
\ee
Using the definition of the twist potential \eqref{twist}, 
\be                       \label{w}
\partial_\rho\chi =\frac{1}{\rho}\,e^{4U}\,\partial_z w,~~~~~\partial_z\chi =-\frac{1}{\rho}\,e^{4U}\,\partial_\rho w,
\ee
these two equations can be represented in the form 
\be
\Delta U+\frac{1}{2}\,e^{-4U}\left[ (\partial_\rho \chi)^2+(\partial_z \chi)^2 \right]&=&0,~~\nn \\
\frac{1}{\rho}\, \partial_\rho\left(\rho\, e^{-4U}\partial_\rho \chi\right)+\partial_z \left(e^{-4U}\partial_z \chi \right)&=&0,
\ee
or, using the compact compact notation, as 
\be            \label{ee}
\Delta U+\frac{1}{2}\,e^{-4U}(\vec{\nabla}\chi)^2=0,~~~~~\vec{\nabla}\left(e^{-4U}\vec{\nabla} \chi\right)=0. 
\ee
Using the Ernst potential \eqref{E}, these two equations can be combined to one  complex-valued equation 
\be           \label{Erns}
(\xi\bar{\xi}-1)\,\Delta\xi=2\bar{\xi}\, (\vec{\nabla}\xi)^2
\ee
usually called in the literature  Ernst equation \cite{Ernst}. However, 
in what follows we shall for simplicity  call  ``Ernst" also equations in the form 
\eqref{ee} or \eqref{Uw}. 
When these equations are solved  and the 
 equation  $\Delta \Phi=0$ is solved as well, 
 the metric function $k$ 
is obtained  from  \eqref{F4}. 
The latter contains two first order equations for $k$ and one second order equation. The first order equations read 
\be             \label{k}
\frac{1}{\rho}\,\partial_\rho k&=&(\partial_\rho U)^2-(\partial_z U)^2
+\frac{1}{4\rho^2}\,e^{4U}[(\partial_z w)^2-(\partial_\rho w)^2]-(\partial_\rho\Phi)^2+(\partial_z\Phi)^2, \nn \\
\frac{1}{2\rho}\,\partial_z k&=&\partial_\rho U \partial_z U 
-\frac{1}{4\rho^2}\,e^{4U}\partial_\rho w\, \partial_z w-\partial_\rho\Phi \partial_z\Phi,
\ee
which can equivalently be rewritten in terms of  the twist $\chi$ instead of rotation $w$. 
The integrability conditions for these equations are insured  
by the Ernst equations and by the equation for the scalar field. 
The second order equation can be represented in the form 
\be              \label{Dk}
\Delta k+2(\partial_z U)^2-2(\partial_z\Phi)^2 +\frac{e^{4U}}{2\rho^2}(\partial_\rho w)^2=0. 
\ee
One can check that this is a differential consequence of the other equations.

%

\section{The dressing procedure of Eris and Gurses -- superposition of solutions }
\setcounter{equation}{0}

The above equations 
split into two independent groups, since the Ernst equation and the scalar field equation are 
independent from each other.  As a result, the solution can be  constructed  in two steps. 
The first step is to consider 
  the purely vacuum problem described by 
the Ernst equation, whose solution determines  $U$, $\chi$ and $w$,
\be
\text{Step I:~~~~~~~~~~Ernst} ~~~~~~\Rightarrow~~~~~U,\chi,w\,.
\ee
This solution is used to compute the amplitude $k_{\rm I}$ defined by  \eqref{k}, where one sets  $\Phi=0$.
This yields explicitly  
\be             \label{kl}
\frac{1}{\rho}\,\partial_\rho k_{\rm I}&=&(\partial_\rho U)^2-(\partial_z U)^2
+\frac{1}{4\rho^2}\,e^{4U}[(\partial_z w)^2-(\partial_\rho w)^2], \nn \\
\frac{1}{2\rho}\,\partial_z k_{\rm I}&=&\partial_\rho U \partial_z U 
-\frac{1}{4\rho^2}\,e^{4U}\partial_\rho w\, \partial_z w. 
\ee
The second step is to solve the scalar field equation,
\be
\text{Step II:~~~~~}\Delta\Phi=0, 
\ee
and compute the amplitude $k_{\rm II}$ from equations obtained from \eqref{k} by keeping there 
only terms with $\Phi$, 
\be             \label{kII}
\frac{1}{\rho}\,\partial_\rho k_{\rm II}=-(\partial_\rho\Phi)^2+(\partial_z\Phi)^2, ~~~~~
\frac{1}{2\rho}\,\partial_z k_{\rm II}=-\partial_\rho\Phi \partial_z\Phi. 
\ee
Taking the sum, 
\be
k=k_{\rm I}+k_{\rm II},
\ee
finally  yields the solution $U,\chi,k,\Phi$ of the equations.  One can say 
that the solution is obtained by superposing (``dressing") a vacuum metric with the scalar field.
This  was first noticed by 
Eris and Gurses  \cite{Eris}. Notice that the dressing only affects the $g_{rr}$ and $g_{\vartheta\vartheta}$ metric 
components while $g_{00}$, $g_{0\varphi}$, $g_{\varphi\varphi}$ are determined by the vacuum equations. 
If one wants the solution to be asymptotically flat, then the scalar field should be bounded, 
but there is only one bounded 
harmonic function, as we shall see. 
Therefore, the scalar field is already known up to a constant factor
and the problem reduces to finding a suitable Ernst potential in the vacuum sector.

\section{Spheroidal coordinates}
\setcounter{equation}{0}

Let us pass from $\rho,z$ to the spheroidal coordinates $x,y=\cos\vartheta$ via 
\be                   \label{sp}
\rho= \sqrt{x^2+\nu}\,\sin\vartheta = \sqrt{(x^2+\nu)(1-y^2)},~~~~z= x\cos\vartheta = x y\,,
\ee
where  $\nu= 0,\pm 1$. One has 
\be
\frac{\rho^2}{x^2+\nu}+\frac{z^2}{x^2}=1\,,
\ee
hence the spheroidal coordinates are oblate if $\nu=1$, prolate if $\nu=-1$, and spherical if $\nu=0$. 
The 3-metric \eqref{h} becomes 
\be                   \label{eK0}
dl^2&=&e^{2k}(d\rho^2+dz^2)+\rho^2 d\varphi^2 \nn \\
&=&e^{2{K}}\left[dx^2+\frac{x^2+\nu}{1-y^2}\,d y^2\right]+(x^2+\nu)(1-y^2) d\varphi^2
\ee
with 
\be                \label{eKa}
e^{2K}=\frac{x^2+\nu y^2}{x^2+\nu}\, e^{2k},
\ee
and the 4-metric is 
\be            \label{eK1}
ds^2=-e^{2U}(dt -w\,d\varphi)^2 +e^{-2U} dl^2.
\ee
Equations \eqref{Uw} assume the form 
\be                     
\label{UU}
&&[(x^2+\nu)U_{,x}]_{,x}+[(1-y^2)U_{,y}]_{,y}+\frac{e^{4U}}{2\rho^2 }\left[
(x^2+\nu) w_{,x}^2+(1-y^2)w_{,y}^2
\right]=0, \nn \\
&&(x^2+\nu)w_{,xx}+(1-y^2)w_{,yy}+4\left[
(x^2+\nu) w_{,x}U_{,x}+(1-y^2)w_{,y}U_{,y}
\right]=0. 
\ee
The relations \eqref{w} between the rotation field and the twist now read 
\be            \label{ww}
w_{,x}=(y^2-1)e^{-4U}\chi_{,y},~~~~w_{,y}=(x^2+\nu)e^{-4U}\chi_{,x}, 
\ee
the integrability condition $\partial_y \chi_{,x}=\partial_x \chi_{,y}$ being insured by the second equation in \eqref{UU}.
Using the twist potential $\chi$ instead of $w$,  Eqs.\eqref{UU} assume the form \eqref{ee},
\be                     
\label{U}
&&[(x^2+\nu)U_{,x}]_{,x}+[(1-y^2)U_{,y}]_{,y}+\frac12\,e^{-4U}\left[
(x^2+\nu) \chi_{,x}^2+(1-y^2)\chi_{,y}^2
\right]=0, \nn \\
&&[(x^2+\nu)\chi_{,x}]_{,x}+[(1-y^2)\chi_{,y}]_{,y}-4\left[
(x^2+\nu) \chi_{,x}U_{,x}+(1-y^2)\chi_{,y}U_{,y}
\right]=0, 
\ee
while the complex Ernst equation \eqref{Erns} becomes 
\be                \label{Er}
(\xi\bar{\xi}-1)\left\{~[(x^2+\nu)\xi_{,x}]_{,x}+[(1-y^2)\xi_{,y}]_{,y}~\right\}=2\bar{\xi}\,\left[
(x^2+\nu)\xi_{,x}^2+(1-y^2)\xi_{,y}^2
\right].
\ee
The scalar field equation reads 
\be                     
\label{Phi}
[(x^2+\nu)\Phi_{,x}]_{,x}+[(1-y^2)\Phi_{,y}]_{,y}=0. 
\ee
Finally, the metric function $K$ defined by \eqref{eKa} can be represented as
\be
K=K_{\rm I}+K_{\rm II}
\ee
where, using \eqref{kl}, $K_{\rm I}$ is defined by 
\be               \label{KK}
\partial_x K_{\rm I}&=&\frac{1-y^2}{x^2+\nu\,y^2}\left( \Gamma(U)
+\frac14\,e^{-4U}\Gamma(\chi)+\frac{\nu\, x}{x^2+\nu}
\right), \nn \\
\partial_y K_{\rm I}&=&\frac{x^2+\nu}{x^2+\nu\,y^2}\left( \Lambda(U)
+\frac14\,e^{-4U}\Lambda(\chi)+\frac{\nu\, y}{x^2+\nu}
\right), 
\ee
or equivalently 
\be               \label{KK00}
\partial_x K_{\rm I}&=&\frac{1-y^2}{x^2+\nu\,y^2}\left( \Gamma(U)
-\frac{e^{4U}}{4\rho^2 }\,\Gamma(w)+\frac{\nu\, x}{x^2+\nu}
\right), \nn \\
\partial_y K_{\rm I}&=&\frac{x^2+\nu}{x^2+\nu\,y^2}\left( \Lambda(U)
-\frac{e^{4U}}{4\rho^2 }\,     \Lambda(w)+\frac{\nu\, y}{x^2+\nu}
\right), 
\ee
with the following definitions 
\be               \label{SL}
\Gamma(f)&\equiv& x(x^2+\nu) f_{,x}^2-2y(x^2+\nu)f_{,x} f_{,y}+x(y^2-1) f_{,y}^2, \nn \\
\Lambda(f)&\equiv& y(x^2+\nu) f_{,x}^2+2x(1-y^2)f_{,x} f_{,y}+y(y^2-1) f_{,y}^2.
\ee
The second part of the amplitude, $K_{\rm II}$,  is defined by Eq.\eqref{kII}, 
\be               \label{KKK}
\partial_x K_{\rm II}&=&-\frac{1-y^2}{x^2+\nu\,y^2}\,\Gamma(\Phi), \nn \\
\partial_y K_{\rm II}&=&-\frac{x^2+\nu}{x^2+\nu\,y^2}\,\Lambda(\Phi).
\ee
A straightforward verification confirms that the condition $\partial_{x}\partial_y K_{\rm I}=\partial_{y}\partial_x K_{\rm I}$ 
is guaranteed by the Ernst equations \eqref{UU},\eqref{U}, while the similar condition for $K_{\rm II}$ 
follows from  the scalar field equation \eqref{Phi}. 
The second order equation \eqref{Dk}  can be represented in the form (after combining it with \eqref{k})
\bea       \label{DK}
&&\left( (x^2+\nu)\,\partial^2_{xx}+x\,\partial_x+(1-y^2)\,\partial^2_{yy}-y\,\partial_y\left)K+\frac{\nu}{x^2+\nu} \right.\right.\\
&&+(x^2+\nu)\left[(\partial_x U)^2-(\partial_x \Phi)^2+\frac{e^{4U}}{4\rho^2}\,(\partial_x w)^2\right] 
+(1-y^2)\left[(\partial_y U)^2-(\partial_y \Phi)^2+\frac{e^{4U}}{4\rho^2}\,(\partial_y w)^2\right]=0. \nn
\eea

As a result, to solve the problem, the 
first step is  to integrate \eqref{UU} or \eqref{U} to find 
$U$ and $w,\chi$ and then compute $K_{\rm I}$ from \eqref{KK} or from \eqref{KK00}. 
This determines the vacuum metric \eqref{eK1} with $K=K_{\rm I}$. 
The second step is to solve the scalar field equation \eqref{Phi} 
and compute the ``dressing"
amplitude $K_{\rm II}$ from \eqref{KKK}. Finally one promotes the vacuum metric to 
the ``dressed" one via replacing $K=K_{\rm I}\to K_{\rm I}+K_{\rm II}$ 
while $U,w$ do not change. The second step of this 
procedure is  essentially trivial, as we shall now see.

\section{Harmonic functions}
\setcounter{equation}{0}

Solutions of the scalar field equation \eqref{Phi} are harmonic functions
\be
\Phi(x,y)=\sum_{l=0}^\infty \,X_l(x) P_l(y),
\ee
where
\be
\left[(x^2+\nu)X_l(x)^\prime\right]^\prime &=& l(l+1)X_l(x),~~~~~\nn \\
\left[(1-y^2)P_l(y)^\prime\right]^\prime &=&-l(l+1)P_l(y). 
\ee
Solutions of the latter equation are the Legendre polynomials, 
$P_0(y)=1$, $P_1(y)=y$, $P_2(y)=3y^2-1$, etc.
Harmonic functions are generically  unbounded, but 
there is one exceptional solution obtained in oblate coordinates, where 
$\nu=1$, in which case one has 
\be
X_0(x)&=&A+C\arctan(x), \nn \\
X_1(x)&=&A\,x+C\,[x\arctan(x)+1], \nn \\
X_2(x)&=&A\,(3x^2+1)+C\,[(3x^2+1)\arctan(x)+3x], \ldots 
\ee
The mode  $X_0(x)$ is  bounded while all the others  are  unbounded. 
For example, one can choose the integration constants $A,C$ such that 
\be               \label{Xpm}
X_1(x)=A^+ X_1^+(x)+A^- X_1^-(x)~~~~\text{with}~~~~X_1^\pm(x)=\frac{x}{2}\pm \frac{1}{\pi}\,(x\arctan(x)+1),
\ee
and when $-\infty\leftarrow x \rightarrow \infty$ one has, respectively, 
\be                  \label{Xpm1}
\frac{1}{3\pi x^2}+\ldots \leftarrow ~~&X_1^+(x)&\rightarrow ~~x+\frac{1}{3\pi x^2}+\ldots ~~~~~\nn \\
x-\frac{1}{3\pi x^2}+\ldots \leftarrow ~~&X_1^-(x)& ~~ \rightarrow -\frac{1}{3\pi x^2}+\ldots
\ee
so that each  $X_1^\pm$ stays finite  either for $x\to\infty$ or for $x\to\-\infty$ but not in both limits. 

If one is interested in globally regular solutions, then the scalar field should be bounded. Therefore, 
the only acceptable solution for the scalar field and the corresponding dressing amplitude defined by \eqref{KKK} are 
\be
\nu=1:~~~~~~~\Phi=C \arctan(x)~~~~\Rightarrow~~~~K_{\rm II}=-\frac{C^2}{2}\,\ln\frac{x^2+y^2}{x^2+1}.       \label{P} 
\ee

\section{Static wormholes}
\setcounter{equation}{0}
Let us see how the dressing  procedure works for static wormholes. 
As a first step, we choose the simplest solution of the Ernst equations \eqref{UU}, 
\be
U=\chi=0,
\ee
in which case 
Eqs.\eqref{KK} yield 
\be           \label{Kr0}
K_{\rm I}=\frac12\,\ln\frac{x^2+\nu y^2}{x^2+\nu}. 
\ee
Setting $\nu=1$ and $K=K_{\rm I}$ yields  the  vacuum metric 
\be                 \label{ring}
ds^2=-dt^2+ \frac{x^2+ y^2}{x^2+1}\,\left[dx^2+\frac{x^2+1}{1-y^2}\,dy^2\right]+(x^2+1)(1-y^2) d\varphi^2.
\ee
Assuming that $x\in (-\infty,\infty)$ and $t={\rm t}/\mu$, $x={\rm r}/\mu$,  this is precisely the ring wormhole \eqref{BE1}.
This metric is locally flat  and the curvature is zero everywhere apart from the conical singularity at the ring $x=y=0$. 
The singularity  is detected by noting that the $x,y$ part of the metric reduces in the vicinity of $x=y=0$    to 
\be                 \label{cone}
(x^2+y^2)(dx^2+dy^2)=r^2(dr^2+r^2d\phi^2)=dR^2+R^2 d\psi^2\,,
\ee
where $x=r\cos\phi$, $y=r\sin\phi$, $R=r^2/2$, $\psi=2\phi$. This is the flat 2D metric in polar coordinates
$R,\psi$, 
however, since $\phi\in[0,2\pi]$,
the angular variable $\psi\in[0,4\pi]$. Therefore, one revolution around $x=y=0$  in the $x,y$ space 
corresponds 
to two revolutions in the  $R,\psi$ space, hence \eqref{cone} is 
the metric on a cone with a negative angle deficit of $-2\pi$. 
This conical singularity can be interpreted as a result of the presence of a 
distributional matter source -- a cosmic string 
of {\it negative} tension extending along the azimuthal $\varphi$-direction  
\cite{Gibbons:2016bok,Gibbons:2017jzk}. In other words, 
this is a loop or  ring made of an infinitely thin  cosmic string. 

Notice that if the range of $x$ was $x\geq 0$, then $x=y=0$ would be at the boundary 
of the $x,y$ space and then one would have  $\phi\in[0,\pi]$ and $\psi\in[0,2\pi]$ in \eqref{cone} 
so that  the conical singularity would be absent. 
Then  \eqref{ring} would be just the Minkowski metric in spheroidal coordinates. 

The presence of a distributional source 
can also be detected by  the equations. The second order equation for $K$ in \eqref{DK}
reduces for $U=w=\Phi=0$ to 
\bea       \label{DK0}
\left( (x^2+1)\,\partial^2_{xx}+x\,\partial_x+(1-y^2)\,\partial^2_{yy}-y\,\partial_y\left)K+\frac{1}{x^2+1} 
=0.\right.\right.~~~~~
\eea
If $K$ is given by \eqref{Kr0} then this equation is apparently fulfilled. However, there is a subtlety 
due to the fact that 
\be
\left( \partial^2_{xx}+\partial^2_{yy}\right)\frac12\ln(x^2+y^2)=2\pi \delta(x)\delta(y).
\ee
This implies that injecting $K$ given  \eqref{Kr0} to \eqref{DK0} does not actually give zero on the right 
but the  delta function instead. This corresponds to a distributional source that should be added 
to the Einstein equations in order that \eqref{ring} be the solution. As a result,  the metric \eqref{ring} 
indeed has a singular source. A more detailed analysis reveals that the 
distributional singularity is contained only in the 
$G_{00}$ and $G_{\varphi\varphi}$ components of the Einstein tensor, 
hence one needs to introduce a source $T_{\mu\nu}$ 
with the only non-vanishing  $T_{00}$ and $T_{\varphi\varphi}$ 
components. 
This corresponds to a cosmic string along the azimuthal 
direction. 

Let us now  add the  scalar field. Choosing the solution \eqref{P} for the scalar,
one has 
\be                  \label{Ksc}
K=K_{\rm I}+K_{\rm II}=\frac{1-C^2}{2}\,\ln\frac{x^2+y^2}{x^2+1}.
\ee
Therefore, 
the $\ln(x^2+y^2)$ term rendering  the metric singular
can be removed by setting 
$C^2=1$, in which case 
\be
K=0,~~~~~\Phi=\pm \arctan(x),
\ee
and the metric becomes 
\be              \label{ring-d}
ds^2=-dt^2+dx^2+\frac{x^2+1}{1-y^2}\,dy^2+(x^2+1)(1-y^2) d\varphi^2.
\ee
This is precisely the ultrastatic BE wormhole \eqref{BE0}. We  obtain  it 
via adding the scalar  field to 
the  vacuum ring wormhole and the scalar screens the singular ring source.
The scalar itself is regular and the screening simply means that 
the resulting geometry with the scalar  is globally regular
and the curvature is everywhere bounded,  so that no extra sources in the equations are needed. 

Two remarks are in order. First, the phantom scalar does not create the wormhole as one might think
but only makes it regular, while the wormhole itself exists already in the vacuum theory. 
Secondly, although the ring source in the static solution seems to be completely screened by the scalar, 
the situation is different in the stationary case, as we shall see below. For stationary solutions 
the scalar field also removes the singularity and makes the geometry regular, but the  memory
of the ring source remains  visible in the metric. Therefore, the regular solutions ``remember"
their descendance from the singular vacuum ring.

The other static BE solutions can be obtained similarly. 
Choosing the complex Ernst potential  $\xi$ to be real and setting 
\be                     \label{BB}
\xi=\tanh(\psi)~~~~~\Rightarrow~~~~{\cal E}=\frac{1-\xi}{1+\xi}=e^{-2\psi}=e^{2U},
\ee
the Ernst equation \eqref{Erns} reduces to \cite{Ernst}
\be
\Delta \psi=0=\Delta U,
\ee
hence the solution is the bounded harmonic function.  In the oblate coordinates, with $\nu=1$, one has, with $\delta$ being 
an integration constant, 
\be
U=\delta\,\arctan(x),~~~~\Rightarrow~~~~~~K_{\rm I}=\frac{\delta^2+1}{2}\,\ln\frac{x^2+y^2}{x^2+1},~
\ee
which determines the vacuum metric with $K=K_{\rm I}$, 
\be                   \label{ZV}
ds^2&=&-e^{2\delta\,\Psi}dt^2+e^{-2\delta\,\Psi} dl^2,\nn \\
dl^2&=& \left(\frac{x^2+ y^2}{x^2+1}\right)^{1+\delta^2}\,\left[dx^2+\frac{x^2+1}{1-y^2}\,dy^2\right]+(x^2+1)(1-y^2) d\varphi^2.
\ee
This is the {\it oblate} ZV metric describing a singular vacuum ring  \cite{Zipoy,Voorhees:1971wh}. 
It reduces to the locally flat metric \eqref{ring} if $\delta=0$. 
The metric singularity at the ring can be removed by adding the scalar dressing \eqref{P} with $C^2=1+\delta^2$, 
which  yields $K=K_{\rm I}+K_{\rm II}=0$.  Therefore, setting 
\be              \label{sol0}
U=\delta\,\Psi,~~~~~\Phi=\pm \sqrt{1+\delta^2}\,\Psi,~~~~~~K=0,
\ee
transforms  \eqref{ZV} to \be
ds^2=-e^{2\delta\,\Psi}dt^2+e^{-2\delta\,\Psi}\left(
\left[dx^2+\frac{x^2+1}{1-y^2}\,dy^2\right]+(x^2+1)(1-y^2) d\varphi^2
\right),
\ee
which coincides with  the regular 
BE metric  \eqref{BE}. 
Finding its  stationary version  requires to find  a spinning 
generalization for the vacuum ZV metric \eqref{ZV} with a subsequent dressing.

 \section{Spinning wormholes -- the relation to the Kerr metric}
 \setcounter{equation}{0}

Let us now start considering stationary generalizations for the wormholes. 
As discussed above, to obtain a spinning version of the ultrastatic BE solution \eqref{BE0}
one has to solve the same problem for the vacuum ring wormhole \eqref{BE1}  
and then add the scalar field. 

What is the stationary version for the ring wormhole ? 
The answer seems to be obvious because, 
as has already been said and will be shown below, the static ring wormhole \eqref{BE1} 
is the special limit of the Kerr metric \cite{Gibbons:2017djb}. Hence 
its stationary generalization  is the Kerr metric itself. Therefore, 
there  remains just  to add the scalar field to the Kerr metric to obtain 
a spinning version of the ultrastatic BE wormhole. 
Let us see, however,  what this gives. 

The Kerr metric is obtained from  the following solution of the Ernst equation \eqref{Er} \cite{Ernst}, 
\be                  \label{Kerr}
\xi=px+iqy~~~~\text{where} ~~~~~q^2-\nu p^2=1~~~\text{with}~~~\nu=\pm 1.
\ee
Reading off   $U,\chi$ from 
\be
{\cal E}=e^{2U}+i\chi=\frac{\xi-1}{\xi+1}
\ee
and computing $w$ and $K_{I}$ from  \eqref{ww}, \eqref{KK}, one obtains 
\bea                \label{xr}
e^{2U}&=&1-\frac{2(px+1)}{(px+1)^2+q^2 y^2},~~~~~~~~\chi=\frac{2qy}{(px+1)^2+q^2 y^2}, \nn \\
w&=&\frac{2q}{p}\times \frac{(px+1)(1-y^2)}{p^2 x^2+q^2 y^2-1},~~~~~
K_{\rm I}=\frac12 \ln\frac{p^2 x^2+q^2 y^2-1 }{x^2+\nu}+K_0. 
\eea
Injecting this to \eqref{eK1} and then to \eqref{mu}, 
setting $K=K_{\rm I}$ and choosing  
\be             \label{xxr}
\mu=\sqrt{\nu\,({\rm a}^2-{\rm M}^2)},~~p=\frac{\mu}{\rm M},~~q=\frac{\rm a}{\rm M},~~K_0=-\ln p, ~~x=\frac{\rm r-M}{\mu}, 
~~t=\frac{\rm t}{\mu},~~y=\cos\vartheta, ~~~~
\ee
yields the Kerr metric in the standard dimensionful form, 
\bea           \label{Kerr-m}
d{\rm s}^2=-d{\rm t}^2 +\frac{\Sigma}{\Delta}\,d{\rm r}^2+\Sigma\, d\vartheta^2 +({\rm r}^2+{\rm a}^2)\sin^2\vartheta\, d\varphi^2 
+\frac{\rm 2Mr}{\Sigma}\left( d{\rm t}- {\rm a}\sin^2\vartheta d\varphi\right)^2, 
\eea
where $\Sigma={\rm r}^2+{\rm a}^2\cos^2\vartheta $ and $\Delta={\rm r^2-2Mr+a^2}$.

It is clear from \eqref{xxr} that $\nu=-1$ corresponds to the ${\rm a<M}$  case when the black hole angular momentum 
is not very high, while $\nu=1$ corresponds to the  supercritical  case when ${\rm a>M}$ and the even horizon is absent
so that the singularity is naked.

As known \cite{Carter:1968rr}, the Kerr metric describes a wormhole geometry with two asymptotic regions 
corresponding to the limits ${\rm r}\to\infty$ and ${\rm r}\to -\infty$. 
Geodesics can interpolate between these two regions, unless
they hit the curvature singularity  located at the ring in the equatorial plane, 
${\rm r}=0$, $\theta=\pi/2$, where one has $\Sigma=0$. 

Taking the ${\rm M}\to 0$ limit with a fixed ${\rm a}$,  which corresponds to the oblate $\nu=1$ regime  and to $\mu=a$, 
the last term in \eqref{Kerr-m} disappears while  the remaining three terms reduce 
exactly to the ring metric \eqref{BE1}.   
The range of the radial coordinate remains the same as for the original Kerr metric, 
hence  the geometry still describes a wormhole but becomes {\it locally} flat (and not flat as often 
stated in the literature). 
As a result, the ${\rm M}\to 0$ limit of the Kerr metric is the static ring wormhole \eqref{BE1}
\cite{Gibbons:2017djb}. Therefore, the natural stationary generalization for   the latter is the 
Kerr metric itself.

It follows that a spinning generalization of the BE wormhole 
will be obtained if we add the scalar dressing to the supercritical Kerr metric. 
Assuming that ${\rm M<a}$ in  \eqref{Kerr-m},  the 
original Papapetrou form of the Kerr metric corresponds to the oblate coordinates, $\nu=1$, 
and then the dressing 
procedure is prescribed  by \eqref{P},
\be
\Phi=C \arctan(x),~~~~~K=K_{\rm I}\to K=K_{\rm I}-\frac{C^2}{2}\,\ln\frac{x^2+y^2}{x^2+1}.       \label{PP} 
\ee
The agreement with the ${\rm M}\to 0$ limit requires that $C^2=1$ and the resulting metric is obtained 
by giving to the $r,\vartheta$ part of the geometry 
\eqref{Kerr-m}
the conformal factor $e^{2K_{\rm II}}=(x^2+1)/( x^2+y^2)$. This  amounts to the replacing in \eqref{Kerr-m}
\be
\frac{\Sigma}{\Delta}\,d{\rm r}^2+\Sigma\, d\vartheta^2~~\to~~ \frac{\Delta}{({\rm r-M})^2+({\rm a^2-M^2})\cos^2\vartheta}
\left(
\frac{\Sigma}{\Delta}\,d{\rm r}^2+\Sigma\, d\vartheta^2
\right). 
\ee
For ${\rm M}=0$ the denominator of the conformal factor cancels against $\Sigma$ and there remains 
$d{\rm r}^2+({\rm r^2+a^2})d\vartheta^2$ hence   the 4-metric reduces  to that for the 
ultrastatic BE wormhole in \eqref{BE0}. However, for ${\rm M}\neq 0$ the denominator
introduces a curvature singularity 
at ${\rm r=M}$, $\vartheta=\pi/2$, in addition to the original 
singularity at ${\rm r}=0$,  $\vartheta=\pi/2$. 
As a result, we do get an exact stationary generalization of the BE solution, but 
it is doubly singular. Therefore, one has to study  other stationary generalizations.

 \section{The relation to the Tomimatsu-Sato metrics}
\setcounter{equation}{0}

This relation is suggested by the following observation. 
The oblate vacuum ZV metric \eqref{ZV} can be obtained 
 by the analytic continuation 
\be                    \label{anal}
x\to ix,~~~~t\to i t,~~~~\delta\to i\delta\,,
\ee
assuming also $\mu\to i\mu$ in \eqref{mu}, 
from the {\it prolate} ZP  metric, 
\be              \label{prol}
ds^2&=&-\left(\frac{x-1}{x+1}\right)^\delta dt^2+
\left(\frac{x-1}{x+1}\right)^{-\delta}
 dl^2,\nn \\
dl^2&=& \left(\frac{x^2- y^2}{x^2-1}\right)^{1-\delta^2}\,\left[dx^2+\frac{x^2-1}{1-y^2}\,dy^2\right]+(x^2-1)(1-y^2) d\varphi^2. 
\ee
This 
reduces to the Schwarzschild metric for $\delta=1$. Its stationary generalizations are explicitly 
known for $\delta=1$ (Kerr metric) and for $\delta=2,3\ldots$. These are the 
 Tomimatsu-Sato (TS) metrics  \cite{Tomimatsu:1972zz,TS} obtained from 
solutions of the vacuum Ernst equation \eqref{Er} in the {\it prolate} ($\nu=-1$) case  
with the complex Ernst potential of the form
\be                \label{Ernst-TS}
\xi_{\rm TS}(p,q,\delta,x,y)=\frac{P(x,y)}{Q(x,y)}\,.
\ee
Here $P,Q$ are polynomials in $x,y$ with coefficients depending on two 
real parameters $p,q$ subject to $p^2+q^2=1$. 
The powers of the polynomials depend on $\delta$, originally assumed to be integer,
but the analysis can be extended to arbitrary real $\delta$ \cite{Cosgrove-I}. 
In the static limit, $q\to 0$, one has 
\be
{\cal E}_{\rm TS}=\frac{\xi_{\rm TS}-1}{\xi_{\rm TS}+1}\to \left(\frac{x-1}{x+1}\right)^\delta~~~~~\text{as}~~~~q\to 0, 
\ee
which corresponds to the 
prolate ZV solution \eqref{prol}.

 This suggests the following 
 procedure: take the stationary TS solution for an arbitrary real $\delta$, then perform the 
  analytic continuation \eqref{anal}, and finally add the scalar dressing. 
 This will give a spinning version of the BE wormhole. The problem, however, is that 
 the TS solution for an arbitrary real $\delta$ is known only in a very implicit form  
 \cite{Cosgrove-I,Cosgrove-II,Cosgrove-III,Cosgrove-IV}, 
\cite{Hori-0,Hori-I,Hori-II,Hori-III,Hori-IV} which does not allow to perform the  analytic continuation. 

Any TS solution for $\nu=-1$ can also be analytically continued via 
\cite{Manko}  
\be              \label{Man}
p\to -i p,~~~~~~x\to ix\,,
\ee
which  yields a solution of the Ernst equation \eqref{Er} for $\nu=+1$. However, 
this continuation is different from \eqref{anal} and does not give what we need. 
 For the 
TS solution with $\delta=0$ the rule \eqref{anal} would reduce just to 
$x\to ix$,  and this  would give a stationary extension for  the ultrastatic  vacuum ring \eqref{BE1}. 
However,  the $\delta=0$ TS solution is not known explicitly either. 

Last but not least,  the ``correct" stationary solution that will be obtained below perturbatively 
does not have the TS form of the Ernst potential. 
Therefore, the TS metrics  are not useful for us, although they do provide some 
stationary solutions for our problem.

\section{Solutions obtained with the harmonic ansatz}
\setcounter{equation}{0}

Exact stationary solutions can be obtained within  the special ansatz which reduces 
the  nonlinear Ernst equations to a single  harmonic equation. 
Choosing the Ernst potential in the form 
\be
\xi=e^{i\alpha}\tanh(\psi),
\ee
one has 
\be
{\cal E}=\frac{1-\xi}{1+\xi}=e^{2U}+i\chi\,,
\ee
where
\be                 \label{UUUU}
e^{-2U}=\cosh(2\psi)+\cos(\alpha)\sinh(2\psi),~~~~~\chi=-\frac{\sin(\alpha)}{\coth(2\psi)+\cos(\alpha)}\,.
\ee
The  Ernst equations \eqref{Erns}
assume the form 
\be               \label{pa}
\Delta \psi-\frac14 (\vec{\nabla}\alpha)^2 \,\sinh(4\psi)&=&0, \nn \\
\Delta\alpha+4\,(\vec{\nabla}\alpha\vec{\nabla}\psi)\,\coth(2\psi)&=&0.
\ee
If $\alpha=const$, these reduce simply to 
\be   \label{lap}
\Delta\psi=0. 
\ee

\subsection{NUT wormholes}

The simplest stationary wormhole, first found  in   \cite{Clement:1983ib}, can be obtained by 
assuming that $\alpha=const$, setting  $\nu=1$, and choosing 
\be                 \label{nut}
\psi=-\delta\,\arctan(x),~~~~~\Phi=\sqrt{\delta^2+1}\,\arctan(x). 
\ee
Injecting this to \eqref{UUUU} and 
computing the  rotation amplitude via \eqref{ww} 
and the $K$-amplitude from \eqref{KK},\eqref{KKK} yields 
\be
w=2\delta\sin(\alpha)\, (y-y_0),~~~~~~K=K_{\rm I}+K_{\rm II}=0. 
\ee
Choosing the integration constant $y_0=1$, the metric is
\be                \label{NUT} 
ds^2=-e^{2U}\left(dt+4\delta\sin(\alpha)\, \sin^2\frac{\vartheta}{2}\, d\varphi\right)^2+e^{-2U}
\left[dx^2+(x^2+1)(d\vartheta^2+\sin^2\vartheta d\varphi^2)\right],~~~~~~
\ee
with $U$ defined by \eqref{UUUU},\eqref{nut}. This stationary solution  reduces to the static 
BE solution \eqref{BE} when $\alpha\to 0$, but for $\alpha\neq 0$ 
it contains the Misner string -- the conical singularity 
along the $\theta=\pi$ axis where $\sin^2({\vartheta}/{2})$ does not vanish. 
The singularity appears because 
$U$ and $\chi$ do not depend on $y$, in which case the rotation field $w$ obtained from \eqref{ww}
is linear in $y=\cos\vartheta$ and  so cannot vanish both for $y=1$ and for $y=-1$. Although 
this singularity is actually quite harmless \cite{Clement:2015cxa,Clement:2015aka},
still its appearance is unpleasant.

\subsection{Removing the NUT singularity}

Still keeping $\alpha=const$, one can avoid the Misner string by letting  $\psi$  depend 
both on $x$ and $y$. In this case the rotation field $w$ is no longer  a liner function of $y$ and one 
can adjust it to vanish both for $y=1$ and for $y=-1$. However, since $y$-depending harmonic functions
are unbounded, the solution will no longer  be asymptotically flat in both limits. 

As the simplest choice, we consider the dipole  mode 
\be           \label{ppp}
\psi={\cal A}  X^{-}_1(x)\, y 
\ee
with $X^{-}_1(x)$ defined by \eqref{Xpm}. Using \eqref{Xpm1}, we see that 
\be
{\cal A} \left( x-\frac{1}{3\pi x^2}+\ldots\right) y ~~\leftarrow~~\psi~~\rightarrow ~~
-\frac{{\cal A} y}{3\pi x^2}+\ldots , 
\ee
and injecting this to \eqref{UUUU} 
it follows  that $U$ ranges in the  limits 
\be
|{\cal A}  y|\, x+\ldots ~~\leftarrow ~~ U~~\rightarrow ~~\frac{{\cal A}\cos\alpha}{3\pi x^2}\, y+\ldots ~~~~~\text{as}~~~~~
-\infty \leftarrow x \rightarrow \infty.
\ee
As $U$ tends to minus infinity for $x\to -\infty$, the geometry is not asymptotically flat in this limit. 
The rotation field is determined from  \eqref{ww}, 
\be          \label{BOT0}
w={\cal A}\sin(\alpha)f(x)(1-y^2)
\ee
with 
\bea           \label{BOT}
f(x)=(x^2+1)\left[\frac12-\frac{1}{\pi}\,\arctan(x)\right]-\frac{x}{\pi}, 
\eea
which ranges within  the limits 
\be          \label{BOT1}
x^2+1+\frac{2}{3\pi x}~~\leftarrow~~ f(x)~~\to~~\frac{2}{3\pi x}+\ldots ~~~~\text{as}~~~~~
-\infty \leftarrow x \rightarrow \infty.
\ee
We see that the rotation field approaches zero for $x\to\infty$ but diverges in the opposite limit. 
At the same time, the $K$-amplitude will be everywhere regular if the 
scalar field is chosen to be the superposition of the dipole and monopole modes,
\be
\Phi=\arctan(x)+{\cal A}  X^{-}_1(x)\, y. 
\ee
Eqs.\eqref{KK},\eqref{KKK} then yield 
\bea
K=K_{\rm I}+K_{\rm II}=\frac{{\cal A}}{\pi}\left[\pi (1-y)+2\arctan(x)\,y-2\arctan\left(\frac{x}{y}\right) \right].
\ee
This function is bounded and ranges in the following limits 
\be          \label{BOT11}
2{\cal A}\,(1-y)+\ldots 
~~\leftarrow~~ K~~\to~~\frac{2{\cal A}\, y(1-y^2)}{3\pi x^3}+\ldots ~~~~\text{as}~~~~~
-\infty \leftarrow x \rightarrow \infty.
\ee
The resulting stationary geometry 
\bea
ds^2=&-&e^{2U}\left( dt - {\cal A}\sin(\alpha)f(x)\sin^2\vartheta\, d\varphi \right)^2 \nn \\
&+& e^{-2U}\left\{
 e^{2K}[dx^2 +(x^2+1)d\vartheta^2]+(x^2+1)\sin^2\vartheta\,d\varphi^2\right\}~~~~~~~
\eea
is free from the  Misner string.  It seems this solution has not been described before.  
The geometry is 
asymptotically flat for $x\to\infty$, with  the angular momentum 
\be
{\rm J}=\frac{{\cal A}\ssigma}{3\pi}\,\sin(\alpha). 
\ee
Curiously, the ADM mass vanishes because $U={\cal O}(1/x^2)$ for $x\to\infty$. 
This solution reduces to the ultrastatic BE wormhole 
when ${\cal A}\to 0$ {\it pointwise} for $x>-\infty$. Therefore, at least when restricted to the $x>-\infty$ region, 
it can be viewed as a rotating generalization for the BE wormhole. However, $U$ and $w$ diverge 
as $x\to-\infty$ hence  the second 
 flat asymptotic is   lost.

\subsection{$\alpha\neq const$}
Let us assume that  $\alpha=\alpha(\psi)$. 
Equations \eqref{pa} then reduce to 
\be                    \label{Del}
\Delta \psi=\frac14\,\alpha^{\prime 2} \,\sinh(4\psi)(\vec{\nabla}\psi)^2, \nn \\
\tanh(2\psi)\,\alpha^{\prime\prime }+\frac12\,\sinh^2(2\psi)\,\alpha^{\prime 3}+4\alpha^\prime =0, 
\ee
which  can be solved  in the parametric form, 
\be             \label{par}
\cosh(2\psi)=\sqrt{1+\eta^2}\,\cosh(Y),~~~~~~\tan(\alpha-\alpha_0)=\eta\,\coth(Y),
\ee
where $\eta,\alpha_0$ are integration constants and $Y$ is a harmonic function, 
$
\Delta Y=0. 
$
This yields a family of new exact stationary solutions. 
The case of constant $\alpha=\alpha_0$ considered above is recovered when $\eta\to 0$. 
Injecting  to \eqref{ww} and defining ${\cal S}=\eta-\sqrt{1+\eta^2}\,\sin(\alpha_0)$
gives 
\be                  \label{WS}
w_{,x}={\cal S}\, (y^2-1)Y_{,y},~~~~~~~~~
w_{,y}={\cal S}\, (x^2+\nu)Y_{,x}. 
\ee
Therefore, 
in order to avoid the Misner string, the harmonic function $Y$ should be $y$-dependent and hence unbounded. 
The $\psi$-amplitude is then also unbounded and 
$U$ in \eqref{UUUU} is  unbounded too. Hence the 
solution  cannot be asymptotically flat in both limits. 

Other exactly solvable cases which similarly reduce to the Laplace equation are 
$U=U(\chi)$ and $S=S(w)$ with $S=\rho^2 e^{-2U}$. 
They always show the same problem -- solutions are not asymptotically flat.

\section{Slowly rotating wormholes}
\setcounter{equation}{0}

One can try and approach the problem differently by assuming the deviation from the static 
limit to be small,  without  restricting the form of the fields. 
Let us  start form the static ring \eqref{BE1} described by the Ernst potential 
\be
\xi=0 
\ee
and try constructing  its slowly rotating version. 
A slowly rotating solution is expected to be a small deformation of the static one,
hence the Ernst potential $\xi$ should be small. Therefore, in the first order of the perturbation theory, it 
should fulfill the linearized Ernst equation \eqref{Erns}, hence 
\be
\Delta \xi=0. 
\ee
As discussed above, the solution must depend both on $x$ and $y$ to avoid the NUT singularity. 
Therefore, it should be unbounded.  This means that  the perturbative approach breaks down, which 
may look like a no-go proof  forbidding  the existence of slowly rotating 
wormholes. Nevertheless, slowly rotating wormholes can be constructed since, in fact,  $\xi=0$ is a 
``wrong vacuum" to expand around. 
As we shall see below, the same static  ring  wormhole can also be described by 
a different solution of the Ernst equation, 
\be
\xi =\tanh\left(\ln(\rho)\right), 
\ee
and the perturbation theory around this  vacuum is well defined. 
We shall see this in the next sections, 
while at the time being let us  see what happens if we
expand around the trivial vacuum $\xi=0$.

Let us choose the  $U,w$ variables and consider the ultrastatic background \eqref{BE1} for which $\nu=1$ and 
\be
U=w=0.
\ee
Small deformations of this solutions are described by 
\be
U=\overset{(1)}{U}+\overset{(2)}{U}+\ldots,~~~~~w=\overset{(1)}{w}+\overset{(2)}{w}+\ldots
\ee
Inserting this to \eqref{Uw} yields in the first order of perturbation theory 
\be
\Delta \overset{(1)}{U}=0,~~~~~
\vec{\nabla}\left( \frac{1}{\rho^2}\, \vec{\nabla} \overset{(1)}{w}  \right)=0, 
\ee
where  one can set $\overset{(1)}{U}=0$, while the $w$-equation explicitly reads 
\be
(x^2+1)\,\overset{(1)}{w}_{,xx}+(1-y^2) \overset{(1)}{w}_{,yy}=0.
\ee
Its solution that vanishes at $y^2=1$ and  is  free from the Misner string is
\be                   \label{w1}
\overset{(1)}{w}={\cal A} f(x) (1-y^2),
\ee
where $f(x)$ 
is the same is in \eqref{BOT}. This solution coincides with that  in \eqref{BOT0}
up to redefining  the integration constant, hence 
it  shows  the same asymptotics which can be written as 
\be                     \label{ww0}
{\cal A}\,\rho^2+\ldots ~~\leftarrow~~ \overset{(1)}{w}~~\rightarrow~~\frac{2{\cal A}\rho^2}{3\pi  x^3}+\ldots
~~~\text{as}~~~~~ -\infty \leftarrow x \to \infty\,,
\ee
where $\rho^2=(x^2+1)\sin^2\vartheta$. 
The solution diverges for $x\to-\infty$, 

In the second order of perturbation theory one has 
\be
\Delta \overset{(2)}{U}+\frac{1}{2\rho^2}\left( \vec{\nabla} \overset{(1)}{w}  \right)^2=0,~~~~~
\vec{\nabla}\left( \frac{1}{\rho^2}\, \vec{\nabla} \overset{(2)}{w}  \right)=0. 
\ee
This is solved by setting $\overset{(2)}{w}=0$ and choosing 
\be
\overset{(2)}{U}=F_0(x)+F_2(x)\,y^2,
\ee
which yields two ODE's for $F_0(x)$ and $F_2(x)$. 
Integrating these equations, the integration constants can be adjusted such that 
$F_0(x)\to 0$ and $F_2(x)\to 0$ for $x\to+\infty$, but 
in the opposite limit these function inevitably diverge, which yields 
\be
-\frac{{\cal A}^2}{4}\,x^2\,(1+y^2)+\ldots \leftarrow \overset{(2)}{U}\to 
\frac{{\cal A}^2}{45\pi x^3}\,(y^2-1) ~~~\text{as}~~~~-\infty \leftarrow x\to \infty.
\ee
We see that  the rotating excitations cannot be small  and diverge for $x\to -\infty$. 
However, there is a different way to carry out the perturbation theory that allows one to 
keep all perturbations finite.

\section{Dualizaton}
\setcounter{equation}{0}

Let $(U,w)$ fulfill  the Ernst equations 
\be                       \label{www}
\Delta U+\frac{e^{4U}}{2\rho^2}(\vec{\nabla} w)^2=0,~~~~~\vec{\nabla}\left( \frac{e^{4U}}{\rho^2}\vec{\nabla} w  \right)=0. 
\ee
The corresponding spacetime metric can be expressed in two different forms, which we shall  call the dual $t$
and $\varphi$ forms: 
\be                \label{p}
{\bm t}:~~~ds^2&=&-e^{2U}(dt-wd\varphi)^2+e^{-2U}\left\{ e^{2 k}(d\rho^2+dz^2) +\rho^2 d\varphi^2\right\}\nn \\
{\bm \varphi}:~~~~~~&=&-\rho^2 e^{-2\OOmega} dt^2+e^{-2\OOmega} e^{2 \kk}(d\rho^2+dz^2) 
+e^{2\OOmega}(d\varphi-\W dt)^2, 
\ee
where 
\be          \label{dual}
e^{2\OOmega}=\rho^2 e^{-2U}-w^2 e^{2U},~~~~~\W e^{2\OOmega}=-w e^{2U},~~~~~\kk-\OOmega=k-U. 
\ee
Notice that $e^{2U}$ is the norm of the timelike Killing vector $\partial/\partial t$
while $e^{2\OOmega}$ is the norm of the azimuthal Killing vector $\partial/\partial\varphi$. In addition, $(\OOmega,\W)$ 
fulfill exactly the same Ernst equations as $(U,w)$ in \eqref{www}, 
\be                \label{Err}
\Delta \OOmega+\frac{e^{4\OOmega}}{2\rho^2}(\vec{\nabla} \W)^2=0,~~~~~\vec{\nabla}\left( \frac{e^{4\OOmega}}{\rho^2}\vec{\nabla} \W \right)=0, 
\ee
while $\kk$ fulfills the same equation as in  \eqref{k}, up to replacing $U\to \OOmega, w\to \W, k\to \kk$. 
As a result, the amplitudes $(\OOmega,\W, \kk)$ 
determine not only the $\varphi$-form of the same solution \eqref{p}, but also 
a new solution with the metric 
\be                \label{pp}
{\bm t}:~~~ds^2&=&-e^{2\OOmega}(dt-\W d\varphi)^2+e^{-2\OOmega}\left\{ e^{2 \kk}(d\rho^2+dz^2) +\rho^2 d\varphi^2\right\}\nn \\
{\bm \varphi}:~~~~~~&=&-\rho^2 e^{-2 U} dt^2+e^{-2 U} e^{2 k}(d\rho^2+dz^2) +e^{2 U}(d\varphi-w \,dt)^2. 
\ee
This solution can formally be obtained from \eqref{p} by the complex change of coordinates
\be
t\to i\varphi,~~~~~~\varphi\to it.
\ee
The inverse transformation $(\OOmega,\W)\to (U,w)$  has exactly the same structure as  \eqref{dual}, 
\be          \label{dual1}
e^{2U}=\rho^2 e^{-2\OOmega}-\W^2 e^{2\OOmega},~~~~~w e^{2U}=-\W e^{2\OOmega},~~~~k-U=\kk-\OOmega.
\ee
Summarizing, solutions of the Ernst equations come in  pairs $(U,w)$ and $(\OOmega,\W)$ related to each other via 
\eqref{dual},\eqref{dual1}. Each pair determines two different geometries \eqref{p} and \eqref{pp}.
  Equivalently, each solution $(U,w)$ of the Ernst equations determines two 
different geometries: either the $t$-geometry defined in \eqref{p} or the $\varphi$-geometry defined in \eqref{pp}. 

Asymptotically flat geometries correspond to solutions $(U,w)$  of the Ernst equations 
for which $e^{2U}\to 1$ at infinity, but also to solutions $(\OOmega,\W)$  
of the Ernst equations  for  which $\rho^2 e^{-2\OOmega}\to 1$
at infinity. 
We have considered above the first option by choosing $U=w=0$ as the background ``vacuum" configuration 
corresponding to the ultrastatic wormhole. However, the same background can be described 
in the dual way by $e^{2\OOmega}=\rho^2$, $\W=0$.

\subsection{Exact solution}

Let us first see if the dual description 
allows one to obtain new exact solutions. 
Introducing the twist potential ${\bm \chi}$ related to $\OOmega,\W$ in the same way as 
in \eqref{w}, one can use for $\OOmega, {\bm \chi}$ 
the same harmonic ansatz as in \eqref{UUUU}. This ansatz expresses the solution  in terms of $\psi,\alpha$
and finally in terms of a  harmonic function $Y$ via \eqref{par}. 
To preserve the asymptotic condition $e^{2\OOmega}\to \rho^2$,
one may choose, for example, 
\be               \label{YYY}
Y=B\ln(\rho)+A\arctan(x), 
\ee
with a suitably adjusted coefficient $B$. 
This yields a family of new  exact stationary solutions. 
However, injecting into \eqref{WS} yields the rotation field 
\be
\W=\left(\eta-\sqrt{1+\eta^2}\,\sin(\alpha_0)\right) (A+2Bx)\,y+\W_0,
\ee
and this  is an unbounded function of $x$ that spoils the asymptotic flatness. 
This function becomes bounded if $B=0$ but then the condition $e^{2\OOmega}\to \rho^2$ is not 
fulfilled hence  the $\varphi$-version of the solution is not asymptotically flat, while its $t$-version 
is similar to \eqref{NUT} and contains the Misner string. 
Adding to \eqref{YYY} extra terms similar to \eqref{ppp} destroys the asymptotic flatness
in both settings. Therefore, one might conclude that the dual formulation does not give anything  
interesting, at least within the harmonic ansatz. 
However, the situation changes if one abandons the ansatz. 

\section{Perturbative analysis in the dual setting} 

Let us start from the metric \eqref{p} in the $\varphi$-form. 
Defining 
\be
e^{2V}=\rho^2 e^{-2\OOmega},~~~~~e^{2{\bm \gamma}}=\rho^2 e^{2{\kk}-4\OOmega},
\ee
the metric becomes 
\bea                \label{pf}
ds^2
=-e^{2V}dt^2+e^{-2V}\left( e^{2 {\bm \gamma}}\,(d\rho^2+dz^2) +\rho^2 (d\varphi-\W dt)^2 \right). 
\eea
The  Ernst equations  \eqref{Err} assume the form 
\be                        \label{EEEx}
\Delta V=\frac{\rho^2}{2} e^{-4V} (\vec{\nabla} \W)^2,~~~~~\vec{\nabla}\left( \rho^2 e^{-4V}\vec{\nabla} \W  \right)=0,
\ee
while Eq.\eqref{k} reduces to 
\be             \label{k-gamma}
\frac{1}{\rho}\,\partial_\rho {\bm \gamma}&=&(\partial_\rho V)^2-(\partial_z V)^2
+\frac{\rho^2}{4}\,e^{-4V}[(\partial_z w)^2-(\partial_\rho w)^2]-(\partial_\rho\Phi)^2+(\partial_z\Phi)^2, \nn \\
\frac{1}{2\rho}\,\partial_z {\bm \gamma}&=&\partial_\rho V \partial_z V 
-\frac{\rho^2}{4}\,e^{-4V}\partial_\rho w\, \partial_z w-\partial_\rho\Phi \partial_z\Phi.
\ee
Passing to the spheroidal coordinates \eqref{sp} yields 
\be               \label{METR}
ds^2
=-e^{2V}dt^2+e^{-2V} \left(e^{2 {\bf {K}}}\left[dx^2+\frac{x^2+\nu}{1-y^2}\, dy^2\right] +(x^2+\nu)(1-y^2) (d\varphi-\W\, dt)^2\right)\,~~~~~~~~~~~
\ee
with 
\be                \label{eK}
e^{2{\bf {K}}}=\frac{x^2+\nu y^2}{x^2+\nu}\, e^{2{\bm\gamma}}. 
\ee
The Ernst  equations \eqref{EEEx} read
\be                     
\label{UUU}
&&[(x^2+\nu)V_{,x}]_{,x}+[(1-y^2)V_{,y}]_{,y}=\frac12\,e^{-4V}(x^2+\nu)(1-y^2)\left[
(x^2+\nu) \W_{,x}^2+(1-y^2)\W_{,y}^2
\right],~~~ \nn \\
&&\frac{[(x^2+\nu)^2\,\W_{,x}]_{,x}}{x^2+\nu}+\frac{[(1-y^2)^2\, \W_{,y}]_{,y}}{1-y^2}=4\left[
(x^2+\nu) \W_{,x}V_{,x}+(1-y^2)\W_{,y}V_{,y}
\right]. ~~
\ee
Setting ${\bf K}={\bf K}_{\rm I}+{\bf K}_{\rm II}$  one obtains from 
Eq.\eqref{k-gamma} 
\be               \label{KK0}
\partial_x {{\bf K}}_{\rm I}&=&\frac{1-y^2}{x^2+\nu\,y^2}\left( \Gamma(V)
-\frac{1}{4}\,(x^2+\nu)(1-y^2) \,e^{-4V}\,\Gamma(\W)+\frac{\nu\, x}{x^2+\nu}
\right), \nn \\
\partial_y {{\bf K}}_{\rm I}&=&\frac{x^2+\nu}{x^2+\nu\,y^2}\left( \Lambda(V)
-\frac{1}{4}\,(x^2+\nu)(1-y^2)\, e^{-4V}\,     \Lambda(\W)+\frac{\nu\, y}{x^2+\nu}
\right), 
\ee
with the same notation as in \eqref{SL}, whereas  ${\bf K}_{\rm II}$ fulfills the same 
equation as $K_{\rm II}$ in \eqref{KKK}, 
\be               \label{KKK11}
\partial_x {\bf K}_{\rm II}=-\frac{1-y^2}{x^2+\nu\,y^2}\,\Gamma(\Phi), ~~~~~
\partial_y {\bf K}_{\rm II}=-\frac{x^2+\nu}{x^2+\nu\,y^2}\,\Lambda(\Phi).
\ee

Let us start from the static solution of \eqref{UUU}, 
\be
\nu=1,~~~~~V=\W=0, 
\ee
which describes either the ultrastatic BE wormhole or the vacuum ring wormhole, 
depending on whether the scalar field is added or not. 
Consider its small deformations, 
\be
V=\overset{(1)}{V}+\overset{(2)}{V}+\ldots,~~~~~\W=\overset{(1)}{\W}+\overset{(2)}{\W}+\ldots
\ee
Inserting this to \eqref{EEEx} yields $\Delta \overset{(1)}{V}=0$,
 whose solution can be chosen to be $\overset{(1)}{V}=0$,
and 
\be
\vec{\nabla}\left( \rho^2\, \vec{\nabla} \overset{(1)}{\W}  \right)=0, 
\ee
which  reads explicitly 
\be            \label{eqW}
\frac{[(x^2+1)^2\,\overset{(1)}{\W}_{,x}]_{,x}}{x^2+1}+\frac{[(1-y^2)^2\, \overset{(1)}{\W}_{,y}]_{,y}}{1-y^2}=0. 
\ee
We remember  that the rotation field $w$ in the $t$-form of the metric should be proportional to 
$\sin^2\vartheta$ to avoid
the NUT singularity. However, 
the same  condition is not needed for the rotation field $\W$ in the $\varphi$-form,
since the $(d\varphi-\W\, dt)^2$ element of the metric \eqref{METR} is  multiplied by 
$1-y^2=\sin^2\vartheta$. 
Therefore, we can assume $\overset{(1)}{\W}$ to depend only on $x$ (otherwise solutions of 
\eqref{eqW} are unbounded), which yields 
\be                   \label{WW}
\overset{(1)}{\W}={\cal A}\,\frac{f(x)}{x^2+1}\equiv 
{\cal A}\,W(x),
\ee
with $f(x)$ defined in \eqref{BOT}. 
When compared  with the previously studied cases \eqref{BOT0}, \eqref{w1}, 
the rotation field now contains an additional factor of $1/(x^2+1)$, hence 
one has 
\be                \label{rot1}
{\cal A}\left(1+\frac{2}{3\pi x^3}+\ldots\right) ~~\leftarrow~~ \overset{(1)}{\W}~~\rightarrow~~{\cal A}\,\frac{2}{3\pi x^3}+\ldots
~~~~\text{as}~~~~-\infty \leftarrow x \rightarrow \infty,
\ee
so that $\overset{(1)}{\W}$ is everywhere bounded and approaches a constant value as $x\to-\infty$. 
This may seem surprizing, since the previously obtained 
perturbative solution \eqref{w1} for the rotation field $\overset{(1)}{w}$ 
was unbounded for $x\to-\infty$, where $\overset{(1)}{w}\sim \rho^2$. 
However, the two results actually agree, because the duality 
transformation \eqref{dual} reduces in the first order of perturbation theory  to 
\be
e^{2\OOmega}=\rho^2 e^{-2U}=e^{2V}=1,~~~~~\overset{(1)}{\W}=-\frac{\overset{(1)}{w}}{\rho^2}.
\ee
Dividing $\overset{(1)}{w}$ in \eqref{w1} by $\rho^2$  yields precisely $\overset{(1)}{\W}$ in \eqref{WW}, 
hence calculations in the $t$-setting and in the $\varphi$-setting  agree. At the same time, the rotation field 
$\overset{(1)}{w}$   is unbounded  whereas $\overset{(1)}{\W}$ 
is bounded, hence the perturbation theory applies in the $\varphi$-setting. 

Now, in the $\varphi$-setting  there is an important  ``twisting" symmetry of the 
line element \eqref{pf},
\be              \label{twis}
\W\to \underline{\W}=\W-\omega,~~~~~\varphi\to \underline{\varphi}= \varphi-\omega\,t\,,
\ee
with a constant $\omega$, which amounts to passing to a rotating frame. 
Applying this with 
$\omega=\overset{(1)}{\W}(-\infty)={\cal A}$ yields the rotation field  in the new frame,  
\be                \label{rot2}
{\cal A}\,\frac{2}{3\pi x^3}+\ldots ~~\leftarrow~~ \overset{(1)}{\underline {\W}}~~\rightarrow~~{\cal A}\,\left(-1+\frac{2}{3\pi x^3}+\ldots\right).
\ee
Comparing with \eqref{rot1}, we see that 
\be             \label{Wr}
\overset{(1)}{\W}(-x)=-\overset{(1)}{\underline{\W}}(x)
\ee 
so that the rotation field is antisymmetric under  
the combined action of the reflection in the wormhole throat, $x\to -x$,
and the  twisting  \eqref{twis}. 
This suggests using  two frames:
the frame where the rotation fulfills \eqref{rot1} should  be used in the $x>0$ region, while the frame 
where the rotation fulfills \eqref{rot2} should be used in the  $x<0$ region. 
Using these two frames, 
the rotation field    approaches zero in both limits, 
for $x\to\infty$ and for $x\to-\infty$.

Let us now continue to the second order of perturbation theory, where 
one can set $\overset{(2)}{\W}=0$, while  
\be                        \label{EEE}
\Delta \overset{(2)}{V}=\frac{\rho^2}{2}  \left(\vec{\nabla} \overset{(1)}{\W}\right)^2,
\ee
or explicitly 
\be                     
\label{UUUa}
&&[(x^2+1)\overset{(2)}{V}_{,x}]_{,x}+[(1-y^2)\overset{(2)}{V}_{,y}]_{,y}=\frac{2{\cal A}^2}{\pi^2}\frac{1-y^2}{(x^2+1)^2}. 
\ee
The variables  can be separated by setting 
\be
\overset{(2)}{V}(x,y)={\cal A}^2\left[\overset{(2)}{V}_{0}(x)+\overset{(2)}{V}_{2}(x)\,y^2\right]. 
\ee
This yields ordinary differential equations for  ${V}_{20}(x)$ and ${V}_{22}(x)$ which admit 
an everywhere bounded solution. The procedure can be continued to 
higher orders of the perturbation theory. In every order the variables can be separated similarly, 
and the integration constants of the ordinary differential equations which appear are
{\it uniquely} fixed by the requirement that the solution should be bounded. As a result, ${\cal A}$ 
is the only integration constant which remains. 
Skipping the details, here is the solution up to 
the forth order terms
\bea                 \label{sol}
\W&=&{\cal A}\, \overset{(1)}w_{0}(x) +{\cal A}^3 \left[\overset{(3)}w_{0}(x) +\overset{(3)}w_{2}(x)\, y^2\right]
+{\cal O}\left({\cal A}^5 \right),   \\
{V}&=&{\cal A}^2\left[\overset{(2)}{V}_{0}(x)+\overset{(2)}{V}_{2}(x)\,y^2\right]+
{\cal A}^4\left[\overset{(4)}{V}_{0}(x)+\overset{(4)}{V}_{2}(x)\,y^2+\overset{(4)}{V}_{4}(x)\,y^4 \right]
+{\cal O}\left({\cal A}^6 \right),\nn 
\eea
where the coefficient functions can be represented as follows,
\bea           \label{pert}
\overset{(1)}w_{0}(x)&=&\frac12(1-W)\,, \nn \\
\overset{(3)}w_{0}(x)&=&-\frac18\,(x^2+1)(W^2-1)W-\frac{x}{3X}\,, \nn \\
\overset{(3)}w_{2}(x)&=&\frac18\,(5x^2+1)(W^2-1)W+\frac{x(6W^2-1)}{3X}+\frac{2W}{X^2}\,, \nn \\
\overset{(2)}{V}_{0}(x)&=&\frac18\,(x^2+1)(W^2-1)\,, \nn \\
\overset{(2)}{V}_{2}(x)&=&-\frac18\,(3x^2+1)(W^2-1)-\frac{xW}{X}-\frac{1}{X^2}\,, \nn \\
\overset{(4)}{V}_{0}(x)&=&\frac{1}{64}\,(x^2+1)^2(W^2+3)(W^2-1)+\frac{xW}{6\pi}\,, \nn \\
\overset{(4)}{V}_{2}(x)&=&-\frac{1}{32}\,(x^2+1)(5x^2+1)(W^2+3)(W^2-1)-\frac{xW(3W^2+7)}{6\pi}
-\frac{3W^2+5}{6\pi X}\,, \nn \\
\overset{(4)}{V}_{4}(x)&=&\frac{1}{192}\,(35\,x^4+30\, x^2+3)(W^2+3)(W^2-1)
+\frac{xW}{18 X}\,[3(5x^2+3)W^2+20x^2+14] \nn \\
&&+\frac{1}{18 X^2}\left[9(3x^2+1)W^2+13x^2+7 \right]
+\frac{4xW}{3X^3}+\frac{2}{3 X^4}\,,
\eea
with the abbreviations $X=\pi\,(x^2+1)$ and 
\be
W=\frac{2}{\pi}\left(\arctan(x)+\frac{x}{x^2+1}\right)=\frac{2}{\pi}\left[x\arctan(x)\right]^\prime\,.
\ee
This function is antisymmetric, $W(x)=-W(-x)$, and one has for $-\infty\leftarrow x\to \infty$
\be
-1-\frac{4}{3\pi x^3}+\ldots ~~\leftarrow~~W(x)~~\rightarrow +1-\frac{4}{3\pi x^3}+\ldots~~~~
\ee
As a result, $\W (-x,y)={\cal A}-\W (x,y)$ and $V(-x,y)=V(x,y)$. 
This yields the lowest terms of the perturbative expansion of the solution of the Ernst equations, and 
higher orders can be included similarly. 
All terms in \eqref{sol} are bounded. One has for $x\to\pm\infty$ 
\bea              \label{as}
V&=&-{\cal A}^2\left( 1+\frac{2}{5}\,{\cal A}^2+\ldots\right)\times \frac{1}{3\pi |x|}+{\cal O}\left(\frac{1}{x^3} \right),\nn  \\
\W&=&\W(\pm\infty)+2{\cal A}\left( 1+\frac{2}{5}\,{\cal A}^2+\ldots\right)\times \frac{1}{3\pi x^3}+{\cal O}\left(\frac{1}{x^4} \right),
\eea
where $\W(\infty)=0$ and  $\W(-\infty)={\cal A}$. 
Computing $e^{2\OOmega}=\rho^2 e^{-2V}$ and injecting together with $\W$ to \eqref{ww} yields 
the twist potential 
\be
{\bm \chi}=\frac{2{\cal A}}{3\pi}\,y\,(3-y^2)&+&{\cal A}^3\, y\left(\frac{4}{15\pi}\,(2y^4-5y^2+5) \right.\nn \\
&+&\left.(1-y^2)^2\left[\frac{X}{4\pi^2}\,(W^2-1)(xXW+2)+\frac{2x^2}{3\pi} \right] \right)+
{\cal O}\left({\cal A}^5\right), 
\ee
from where one can see that the complex Ernst potential 
$e^{2\OOmega}+i{\bm \chi}=\rho^2 e^{-2 V}+i{\bm \chi}$ 
is not of  the Tomimatsu-Sato type since it contains powers of $x,y$ and also of $\arctan(x)$. 

To complete the line element, there remains to determine the ${\bf K}$-amplitude. 
Integrating \eqref{KK0} yields 
\be
{\bf K}_{\rm I}={\bf K}_{\rm reg}+{\bf K}_{\rm sing}\,,
\ee
where the regular part is 
\bea              \label{Kreg}
{\bf K}_{\rm reg}= && \left({\cal A}^2+ {\cal A}^4\left[
\frac13(2-y^2)+y^2W^2 +y^2 xXW(W^2-1)
 \right.\right.  \\
+
&&\left.\left.
\frac{X^2}{32}[y^2-1+x^2(9y^2-1) ](W^2-1)^2
 \right]
+{\cal O}\left({\cal A}^6 \right)\right)
\times \frac{1}{2\pi^2}\,\frac{y^2-1}{x^2+1}\,,
 \nn
\eea
and the singular part 
\be              \label{sing} 
{\bf K}_{\rm sing}=\left(
1-\frac{{\cal A}^2}{\pi^2}+\frac{{\cal A}^4}{3\pi^2}+{\cal O}\left({\cal A}^6 \right)
\right)
\times \frac12\,\ln\frac{x^2+y^2}{x^2+1}\,.
\ee
If we assume that $\Phi=0$ then ${\bf K}={\bf K}_{\rm I}$ and the above solution describes
the spinning  generalization of the vacuum ring wormhole. If ${\cal A}=0$, then 
the 4-metric \eqref{METR} becomes flat everywhere away  from the circle $x=y=0$
where the distributional singularity of the Ricci tensor is located. As explained  above, this singularity can be interpreted 
as a singular matter source  made  of a ring of negative tension. 
For ${\cal A}\neq 0$ the ring rotates in the equatorial plane, the geometry is then 
 no longer locally flat and its singularity has also a volume part 
containing components of the Riemann tensor which diverge as one approaches the ring. 
However, the solution remains a wormhole with two asymptotically flat regions, and the 
geodesics can interpolate between these regions, unless they hit the ring singularity. 

Let us finally add the scalar field according to \eqref{P}, 
\be
\Phi=C\,\arctan(x)~~~\Rightarrow~~~~{\bf K}_{\rm II}=-\frac{C^2}{2}\ln \frac{x^2+y^2}{x^2+1}.
\ee
Comparing with \eqref{sing}, 
we see that ${\bf K}_{\rm II}+{\bf K}_{\rm sing}=0$ 
and hence 
the metric singularity contained in ${\bf K}_{\rm sing}$ is cancelled if 
the integration constant  is chosen as 
\be
C^2 = 1-\frac{{\cal A}^2}{\pi^2}+\frac{{\cal A}^4}{3\pi^2}+{\cal O}\left({\cal A}^6 \right).
\ee
One has then ${\bf K}={\bf K}_{\rm reg}$ and the spacetime geometry 
becomes everywhere regular because all three metric functions $V,\W,{\bf K}$ in the lime element 
\eqref{METR} are bounded. 
One has 
for $x\to \pm\infty$ 
\be
{\bf K}={\bf K}_{\rm reg}= \left({\cal A}^2+\frac49\,{\cal A}^4+{\cal O}\left({\cal A}^6 \right)\right)\times \frac{y^2-1}{2\pi x^2}
+{\cal O}\left(\frac{1}{x^3} \right). 
\ee
 One can say that  the singular ring source is screened by the scalar, 
which  yields the globally regular spinning generalization for the ultrastatic BE wormhole. 
Its curvature is everywhere bounded and approaches zero in the two asymptotic regions. 

At the same time, 
the metric components 
\be            \label{g00}
g_{00}=-e^{2U}=-e^{2V}+\rho^2 \W^2\,e^{-2V},~~~~~~~~ g_{0\varphi}=-\rho^2 \W e^{-2V}
\ee
are unbounded since for $-\infty\leftarrow x\to \infty$ one has 
\be
\W^2(-\infty)\times  \rho^2 ~~\leftarrow ~~g_{00}~~\to~~ -1,~~~~~~
-\W(-\infty)\times  \rho^2 ~~\leftarrow ~~g_{0\varphi}~~\to~~ 0.
\ee
The spacetime contains an ergoregion where $g_{00}=-e^{2U}$ becomes positive and  the timelike 
Killing vector becomes spacelike. Therefore, the Newtonian potential  $U$ is not globally defined. However, there exists 
a linear combination of  the  two Killing vectors which remains timelike in the ergoregion, while 
the metric components  can be made finite for $x<0$ by the 
twisting transformation \eqref{twis},\eqref{rot2}. This changes the frame and the asymptotic value of $\W$. 
Let us define two  rotation fields  $\W_{+}=\W$ and 
$\W_{-}=\W-\W(-\infty)$ such that 
for $-\infty\leftarrow x\to \infty$ one has 
\be
{\cal A}~~\leftarrow~~\W_{+}~~\to ~~0,~~~~~~~~~
0~~\leftarrow~~\W_{-}~~\to ~~-{\cal A}.
\ee
One should use $\W_{+}$ to compute the metric components $g_{00}$ and $g_{0\varphi}$ in the 
$x>0$ region and one uses $\W_{-}$ to compute them in the $x<0$ region. 
This allows one to compute  the  ADM mass $M$ and angular momentum $J$.
These are the same for the ring wormhole and for the BE wormhole, since 
 $g_{00}$ and $g_{0\varphi}$ are the same. 
One has for $x\to\pm\infty$ 
\bea           \label{gg}
-g_{00}&=&e^{2V}-\rho^2 \W_\pm^2 e^{-2V}=1-\frac{2M_{\pm }}{|x|}+\ldots ,\nn \\
-g_{0\varphi}&=&\rho^2 \W_\pm \, e^{-2V}=\frac{2J_{\pm }\sin^2\vartheta }{|x|}+\ldots 
\eea
This determines the mass $M_\pm$ and angular momentum $J_\pm$ 
measured, respectively, at $x\to\pm\infty$. Notice that the denominators in \eqref{gg} should contain 
$|x|$ and not $x$ since the mass and angular momentum should be invariant 
under the coordinate transformation $x\to -x$. 

It is important to emphasise that $g_{00}$ and $g_{0\varphi}$ and hence $M_\pm$ and $J_\pm$ 
are determined by the vacuum equations and are insensitive  to the scalar field. Including the latter 
only modifies the ${\bf K}$-amplitude without affecting $g_{00}$ and $g_{0\varphi}$. 

Using \eqref{as} in \eqref{gg} and 
restoring the length scale  gives the 
dimensionful values, 
\be                    \label{ADM}
{\rm M}\equiv {\rm M}_\pm =\frac{\mu}{3\pi}{\cal A}^2\left( 1+\frac{2}{5}\,{\cal A}^2+\ldots\right),~~~~~
{\rm J}_\pm=\pm \frac{\mu}{3\pi}{\cal A}\left( 1+\frac{2}{5}\,{\cal A}^2+\ldots\right),
\ee
so that the mass is the same and positive in each asymptotic region, 
while the  angular momentum changes sign 
when one passes from one region to the other one. The latter property is clear, since if the hole 
rotates in the clockwise direction, say, when viewed from one asymptotic region, then 
it rotates in the opposite 
direction when viewed from the other region. 

One can establish an exact Smarr-type relation between $M$ and $J$ \cite{Kleihaus:2014dla}. 
The two Ernst equations \eqref{EEEx} can be combined to yield 
\be
\Delta V=\vec{\nabla}
\left(\frac{\rho^2}{2}\,e^{-4V}\W\vec{\nabla}\W\right).
\ee
Integrating this over $x$ from $x_1$ to $x_2$ and over $y$ from $-1$ to $1$ one obtains 
\be
\left. \int_{-1}^{1} dy\, (x^2+1)\, \partial_x V\right|_{x_{1}}^{x_{2}}=
\frac12 \left. \int_{-1}^{1} dy\,(1-y^2) (x^2+1)^2\,e^{-4V} \W\partial_x \W \right|_{x_{1}}^{x_{2}}. 
\ee
Choosing first  $x_1=0$, $x_2=\infty$ and next $x_1=-\infty$, $x_2=0$, using
\eqref{as} and the fact that $V$ is a symmetric function of $x$, it is not difficult to see that 
\be          \label{Smarr}
{\rm M}={\cal A}\,{\rm J}_{+}\,.
\ee
Since $\W(-x,y)={\cal A}-\W(x,y)$, it follows that $\W(0,y)={\cal A}/2\equiv \W^0$
hence the rotation field assumes a constant value  in the wormhole throat -- 
the throat angular velocity. This depends on the frame. For the two 
rotation fields $\W_{+}=\W$ and $\W_{-}$ 
one has $\W^0_\pm=\pm {\cal A}/2$, which allows one to represent \eqref{Smarr} as 
\be          \label{Smarr1}
{\rm M}=2\,\W^0_{\pm}\,{\rm J}_{\pm}. 
\ee
This, however, does not mean that $M$ is a linear function of $J$, since the proportionality coefficient 
is $J$-dependent. 
As seen from \eqref{ADM}, one has for small $J$
\be                 \label{Vir}
{\rm M}=\frac{3\pi}{\mu} {\rm J}^2+{\cal O}({\rm J}^4).
\ee
This  is worth comparing with the expression for the rotational energy of a non-relativistic 
rigid body,
\be
{\rm E}_{\rm rot}=\frac{\rm J^2}{2I},
\ee
where $I$ is the moment of inertia. Such a relation is extected   for the 
ring wormhole, since it contains the extended matter source -- the cosmic string loop.
For a slow rotation, this spinning string should exhibit the standard non-relativistic relation. 
It is however quite remarkable that 
the spinning BE wormhole also shows exactly the same relation \eqref{Vir}, because it has the same ${\rm M}$ 
and ${\rm J}$ as the ring wormhole.  Although it is globally regular, it 
 ``remembers" its descendance from the ring wormhole. 

This  observation is important. 
Although the BE wormhole contains the scalar field, the latter  is spherically symmetric and 
does not carry rotational degrees of freedom.
The rotation is encoded in the $g_{00}$ and  $g_{0\varphi}$ metric components, which are 
determined by the vacuum theory with the ring source and are insensitive to the scalar.  
The scalar only modifies the $g_{rr}$ and $g_{\vartheta\vartheta}$ components 
to hide the source and make the 
geometry regular. However, the source is still visible in $g_{00}$, $g_{0\varphi}$
 and the BE wormhole ``knows"  about it. 
Therefore, the essential features -- the wormhole structure itself and the rotation -- 
originate in the vacuum theory and exist due to the 
ring source and not due to the phantom scalar field as one might  have thought. The
only role of the 
 scalar  is to render 
the geometry regular.

\section{Concluding remarks  -- toward the exact solution ?}

Summarizing, we have described a number of possible ways to construct  the  stationary 
generalization for the  static BE wormholes supported by the phantom scalar field.  
Perhaps not obviously  important physically, since the BE wormhole is  unstable \cite{Shinkai:2002gv}, 
this problem is interesting  conceptually, because it is important to understand why it is so difficult to obtain the 
stationary version of the static solution \eqref{BE0} which looks  much simpler than the 
Schwarzschild solution. 

We find that the difficulty is actually not related to the scalar field,  which  can 
be eliminated  within  the Eris-Gurses procedure.  The problem reduces to constructing  the  stationary 
generalization for  the vacuum ring wormhole via 
solving the vacuum Ernst equations, and it is the latter step which is difficult. 
Even though the static wormhole geometry \eqref{BE1}  is {\it locally flat}, its stationary generalization  is difficult to obtain 
and it is of  a previously  unknown type.

Using the special ansatz to solve the Ernst equations, we have constructed exact solutions, but they 
are not globally regular.  The perturbative expansion 
around the trivial solution of the Ernst equation
which describes  the static limit,  ${\cal E}=e^{2U}+i\chi=1$,   contains unbounded functions and is ill-defined.  
However, the static limit can also be described  by 
$e^{2\OOmega}+i{\bm \chi}=\rho^2$ where  $e^{2\OOmega}$ and ${\bm \chi}$ are the norm and twist 
of the axial Killing vector.  The perturbative expansion around this vacuum 
is described by \eqref{sol} and contains only bounded functions.
Although not a proof, this gives  a good indication for the existence of 
a fully non-perturbative solution. An additional   indication is provided 
by the the numerical analysis in \cite{Kleihaus:2014dla,Chew:2016epf}, 
which shows  a numerical solution whose properties seem to correspond to our solution. 
This seems to be  the same solution, which 
gives an extra  evidence in favour of its existence.

The solution describes the spinning generalization for the locally flat vacuum ring 
wormhole and that for the ultrastatic BE wormhole, depending on whether the scalar field is included or not. 
The spinning wormhole interpolates between two asymptotically flat regions  and is characterised by a non-zero ADM 
mass proportional to the square of the angular momentum, which is typical for a rotating extended source. 
The ring wormhole shows the ring singularity but the spinning BE wormhole is globally regular. 
Apart from this difference, the singular and regular solutions have identically the same 
$g_{00}$ and $g_{0\varphi}$ metric components and the same ADM mass and angular momentum 
determined 
by the  ring source and not by the phantom scalar field as one might have expected. 
The only role of the scalar is to screen the metric singularity  at the ring source 
and make the geometry globally regular, but the memory  of the source remains in $g_{00}$ and $g_{0\varphi}$.

At the same time, our initial intension  was to obtain the solution exactly.  Therefore,
there remains the  question of 
weather the perturbative solution  \eqref{sol} could  be 
promoted to an exact one. However, since the expansion contains  powers of $x$ and $y$
and also of $\arctan(x)$, there is 
little hope to guess the  exact form of the solution, while the other  known methods to get the solution
do not seem to work. 
For example, it is known that the Ernst equations are equivalent to one 
fourth order PDE for the metric function $k$  \cite{Cosgrove-III}. For  the Tomimatsu-Sato solutions, 
as for example 
for the Kerr metric in \eqref{xr}, one always has $k(x,y)=k(\eta)$ with 
$$
\eta=\frac{x^2+\nu y^2}{x^2+\nu}. 
$$
This implies that the fourth order equation for $k$ actually becomes an ODE,  which 
allows one to obtain exact solutions \cite{Cosgrove-I,Cosgrove-II,Cosgrove-III,Cosgrove-IV}. 
However, neither the  function ${\bf K}$ in \eqref{Kreg},\eqref{sing} nor 
the amplitude $k$ or its $\varphi$-counterpart ${\bf k}$ defined by \eqref{p} 
expressed by 
$$
k={\bf K}+\ln\left(1-\rho^2\W^2 e^{-4V} \right)-\frac12\ln(\eta),~~~~~~~
{\bf k}={\bf K}-2V+\frac12\ln\frac{\rho^2}{\eta}$$
depend   exclusively  on $\eta$. Therefore, neither $k$ not ${\bf k}$ 
satisfy an ODE, hence  this approach does not allow to get  the  solution exactly.

 The static ring wormholes described by the oblate ZV metrics can be 
 promoted to the stationary sector 
  by applying the solution generating methods, 
 but this yields non-asymptotically flat solutions with an electric field 
\cite{Clement:1997tx,Clement:1998nk}. It is also not obvious  if   the 
 inverse scattering method \cite{Belinski:2001ph} could be helpful, 
 although this possibility deserves a separate study.

 The reason for the difficulties in finding the exact solution is  clear. 
 The analytically known stationary metrics 
  like Kerr describe  spinning states 
  of zero-dimensional objects --  massive points. However, the static vacuum geometry \eqref{BE1} 
 has an extended one-dimensional  source:  the 
 ring. Therefore,  constructing  its stationary   version should be  a more complex problem 
 that may not have an analytical solution. 
 
 \acknowledgements
It is a pleasure to thank Gary Gibbons for discussions. 
This work was partly supported by the French  CNRS/RFBR PRC grant No.289860
and also by the Russian Foundation for Basic Research on the project 20-52-18012.


\begin{thebibliography}{10}

\bibitem{Einstein:1935tc}
A.~Einstein and N.~Rosen, {\it {The particle problem in the General Theory of
  Relativity}},  {\sl Phys.Rev.} {\bf 48} (1935) 73--77,
  [\href{http://dx.doi.org/10.1103/PhysRev.48.73}{{\sf
  doi:10.1103/PhysRev.48.73}}].

\bibitem{Visser:1995cc}
M.~Visser, {\em {Lorentzian wormholes: From Einstein to Hawking}}.
\newblock AIP, 1996.

\bibitem{Friedman:1993ty}
J.~L. Friedman, K.~Schleich, and D.~M. Witt, {\it {Topological censorship}},
  {\sl Phys.Rev.Lett.} {\bf 71} (1993) 1486--1489,
  [\href{http://arxiv.org/abs/gr-qc/9305017}{{\sf arXiv:gr-qc/9305017}}],
  [\href{http://dx.doi.org/10.1103/PhysRevLett.71.1486}{{\sf
  doi:10.1103/PhysRevLett.71.1486}}]. [Erratum: Phys. Rev. Lett.75,1872(1995)].

\bibitem{Hochberg:1998ii}
D.~Hochberg and M.~Visser, {\it {The null energy condition in dynamic
  wormholes}},  {\sl Phys.Rev.Lett.} {\bf 81} (1998) 746--749,
  [\href{http://arxiv.org/abs/gr-qc/9802048}{{\sf arXiv:gr-qc/9802048}}],
  [\href{http://dx.doi.org/10.1103/PhysRevLett.81.746}{{\sf
  doi:10.1103/PhysRevLett.81.746}}].

\bibitem{Morris:1988tu}
M.~Morris, K.~Thorne, and U.~Yurtsever, {\it {Wormholes, time machines, and the
  weak energy condition}},  {\sl Phys.Rev.Lett.} {\bf 61} (1988) 1446--1449,
  [\href{http://dx.doi.org/10.1103/PhysRevLett.61.1446}{{\sf
  doi:10.1103/PhysRevLett.61.1446}}].

\bibitem{Bronnikov:1973fh}
K.~Bronnikov, {\it {Scalar-tensor theory and scalar charge}},  {\sl Acta
  Phys.Polon.} {\bf B4} (1973) 251--266.

\bibitem{Ellis:1973yv}
H.~G. Ellis, {\it {Ether flow through a drainhole - a particle model in general
  relativity}},  {\sl J.Math.Phys.} {\bf 14} (1973) 104--118,
  [\href{http://dx.doi.org/10.1063/1.1666161}{{\sf doi:10.1063/1.1666161}}].

\bibitem{1538-3881-116-3-1009}
A.~Riess {\em et~al.}, {\it Observational evidence from supernovae for an
  accelerating universe and a cosmological constant},  {\sl Astron.Journ.} {\bf
  116} (1998), no.~3 1009.

\bibitem{0004-637X-517-2-565}
S.~Perlmutter {\em et~al.}, {\it Measurements of $\omega$ and $\lambda$ from 42
  high-redshift supernovae},  {\sl Astrophys.Journ.} {\bf 517} (1999), no.~2
  565.

\bibitem{Kanti:2011jz}
P.~Kanti, B.~Kleihaus, and J.~Kunz, {\it {Wormholes in Dilatonic
  Einstein-Gauss-Bonnet Theory}},  {\sl Phys.Rev.Lett.} {\bf 107} (2011)
  271101, [\href{http://arxiv.org/abs/1108.3003}{{\sf arXiv:1108.3003}}],
  [\href{http://dx.doi.org/10.1103/PhysRevLett.107.271101}{{\sf
  doi:10.1103/PhysRevLett.107.271101}}].

\bibitem{Cuyubamba:2018jdl}
M.~A. Cuyubamba, R.~A. Konoplya, and A.~Zhidenko, {\it {No stable wormholes in
  Einstein-dilaton-Gauss-Bonnet theory}},  {\sl Phys. Rev. D} {\bf 98} (2018),
  no.~4 044040, [\href{http://arxiv.org/abs/1804.11170}{{\sf
  arXiv:1804.11170}}],
  [\href{http://dx.doi.org/10.1103/PhysRevD.98.044040}{{\sf
  doi:10.1103/PhysRevD.98.044040}}].

\bibitem{Bronnikov:2002rn}
K.~Bronnikov and S.-W. Kim, {\it {Possible wormholes in a brane world}},  {\sl
  Phys.Rev.} {\bf D67} (2003) 064027,
  [\href{http://arxiv.org/abs/gr-qc/0212112}{{\sf arXiv:gr-qc/0212112}}],
  [\href{http://dx.doi.org/10.1103/PhysRevD.67.064027}{{\sf
  doi:10.1103/PhysRevD.67.064027}}].

\bibitem{Sushkov:2011jh}
S.~V. Sushkov and R.~Korolev, {\it {Scalar wormholes with nonminimal derivative
  coupling}},  {\sl Class.Quant.Grav.} {\bf 29} (2012) 085008,
  [\href{http://arxiv.org/abs/1111.3415}{{\sf arXiv:1111.3415}}],
  [\href{http://dx.doi.org/10.1088/0264-9381/29/8/085008}{{\sf
  doi:10.1088/0264-9381/29/8/085008}}].

\bibitem{Sushkov:2015fma}
S.~V. Sushkov and M.~S. Volkov, {\it {Giant wormholes in ghost-free bigravity
  theory}},  {\sl JCAP} {\bf 1506} (2015), no.~06 017,
  [\href{http://arxiv.org/abs/1502.03712}{{\sf arXiv:1502.03712}}],
  [\href{http://dx.doi.org/10.1088/1475-7516/2015/06/017}{{\sf
  doi:10.1088/1475-7516/2015/06/017}}].

\bibitem{Clement:1983ic}
G.~Clement, {\it {Regular multiparticle solutions of Einstein-Maxwell scalar
  field theories}},  {\sl Class.Quant.Grav.} {\bf 1} (1984) 275,
  [\href{http://dx.doi.org/10.1088/0264-9381/1/3/006}{{\sf
  doi:10.1088/0264-9381/1/3/006}}].

\bibitem{Clement:2015lul}
G.~Clement, {\it {Axisymmetric multiwormholes revisited}},  {\sl Gen.Rel.Grav.}
  {\bf 48} (2016), no.~6 76, [\href{http://arxiv.org/abs/1511.06249}{{\sf
  arXiv:1511.06249}}], [\href{http://dx.doi.org/10.1007/s10714-016-2073-y}{{\sf
  doi:10.1007/s10714-016-2073-y}}].

\bibitem{Egorov:2016rfr}
A.~I. Egorov, P.~E. Kashargin, and S.~V. Sushkov, {\it {Scalar
  multi-wormholes}},  {\sl Class.Quant.Grav.} {\bf 33} (2016), no.~17 175011,
  [\href{http://arxiv.org/abs/1603.09552}{{\sf arXiv:1603.09552}}],
  [\href{http://dx.doi.org/10.1088/0264-9381/33/17/175011}{{\sf
  doi:10.1088/0264-9381/33/17/175011}}].

\bibitem{Gibbons:2017jzk}
G.~W. Gibbons and M.~S. Volkov, {\it {Weyl metrics and wormholes}},  {\sl JCAP}
  {\bf 1705} (2017), no.~05 039, [\href{http://arxiv.org/abs/1701.05533}{{\sf
  arXiv:1701.05533}}],
  [\href{http://dx.doi.org/10.1088/1475-7516/2017/05/039}{{\sf
  doi:10.1088/1475-7516/2017/05/039}}].

\bibitem{Gibbons:2016bok}
G.~W. Gibbons and M.~S. Volkov, {\it {Ring wormholes via duality rotations}},
  {\sl Phys.Lett.} {\bf B760} (2016) 324--328,
  [\href{http://arxiv.org/abs/1606.04879}{{\sf arXiv:1606.04879}}],
  [\href{http://dx.doi.org/10.1016/j.physletb.2016.07.012}{{\sf
  doi:10.1016/j.physletb.2016.07.012}}].

\bibitem{Yazadjiev:2017twg}
S.~Yazadjiev, {\it {Uniqueness theorem for static wormholes in Einstein-phantom
  scalar field theory}},  {\sl Phys. Rev. D} {\bf 96} (2017), no.~4 044045,
  [\href{http://arxiv.org/abs/1707.03654}{{\sf arXiv:1707.03654}}],
  [\href{http://dx.doi.org/10.1103/PhysRevD.96.044045}{{\sf
  doi:10.1103/PhysRevD.96.044045}}].

\bibitem{Deligianni:2021ecz}
E.~Deligianni, J.~Kunz, P.~Nedkova, S.~Yazadjiev, and R.~Zheleva, {\it
  {Quasiperiodic oscillations around rotating traversable wormholes}},  {\sl
  Phys. Rev. D} {\bf 104} (2021), no.~2 024048,
  [\href{http://arxiv.org/abs/2103.13504}{{\sf arXiv:2103.13504}}],
  [\href{http://dx.doi.org/10.1103/PhysRevD.104.024048}{{\sf
  doi:10.1103/PhysRevD.104.024048}}].

\bibitem{Teo:1998dp}
E.~Teo, {\it {Rotating traversable wormholes}},  {\sl Phys. Rev. D} {\bf 58}
  (1998) 024014, [\href{http://arxiv.org/abs/gr-qc/9803098}{{\sf
  arXiv:gr-qc/9803098}}],
  [\href{http://dx.doi.org/10.1103/PhysRevD.58.024014}{{\sf
  doi:10.1103/PhysRevD.58.024014}}].

\bibitem{Clement:1983ib}
G.~Clement, {\it {A class of stationary axisymmetric solutions of
  Einstein-Maxwell scalar field theories}},  {\sl Class.Quant.Grav.} {\bf 1}
  (1984) 283, [\href{http://dx.doi.org/10.1088/0264-9381/1/3/007}{{\sf
  doi:10.1088/0264-9381/1/3/007}}].

\bibitem{Matos:2009au}
T.~Matos, {\it {Class of Einstein-Maxwell phantom fields: rotating and
  magnetised wormholes}},  {\sl Gen. Rel. Grav.} {\bf 42} (2010) 1969--1990,
  [\href{http://arxiv.org/abs/0902.4439}{{\sf arXiv:0902.4439}}],
  [\href{http://dx.doi.org/10.1007/s10714-010-0976-6}{{\sf
  doi:10.1007/s10714-010-0976-6}}].

\bibitem{Kashargin:2007mm}
P.~E. Kashargin and S.~V. Sushkov, {\it {Slowly rotating wormholes: The First
  order approximation}},  {\sl Grav. Cosmol.} {\bf 14} (2008) 80--85,
  [\href{http://arxiv.org/abs/0710.5656}{{\sf arXiv:0710.5656}}],
  [\href{http://dx.doi.org/10.1134/S0202289308010106}{{\sf
  doi:10.1134/S0202289308010106}}].

\bibitem{Kashargin:2008pk}
P.~E. Kashargin and S.~V. Sushkov, {\it {Slowly rotating scalar field
  wormholes: The second order approximation}},  {\sl Phys. Rev. D} {\bf 78}
  (2008) 064071, [\href{http://arxiv.org/abs/0809.1923}{{\sf
  arXiv:0809.1923}}], [\href{http://dx.doi.org/10.1103/PhysRevD.78.064071}{{\sf
  doi:10.1103/PhysRevD.78.064071}}].

\bibitem{Kleihaus:2014dla}
B.~Kleihaus and J.~Kunz, {\it {Rotating Ellis wormholes in four dimensions}},
  {\sl Phys. Rev. D} {\bf 90} (2014) 121503,
  [\href{http://arxiv.org/abs/1409.1503}{{\sf arXiv:1409.1503}}],
  [\href{http://dx.doi.org/10.1103/PhysRevD.90.121503}{{\sf
  doi:10.1103/PhysRevD.90.121503}}].

\bibitem{Chew:2016epf}
X.~Y. Chew, B.~Kleihaus, and J.~Kunz, {\it {Geometry of spinning Ellis
  wormholes}},  {\sl Phys. Rev. D} {\bf 94} (2016), no.~10 104031,
  [\href{http://arxiv.org/abs/1608.05253}{{\sf arXiv:1608.05253}}],
  [\href{http://dx.doi.org/10.1103/PhysRevD.94.104031}{{\sf
  doi:10.1103/PhysRevD.94.104031}}].

\bibitem{Clement:1997tx}
G.~Clement, {\it {From Schwarzschild to Kerr: Generating spinning
  Einstein-Maxwell fields from static fields}},  {\sl Phys. Rev. D} {\bf 57}
  (1998) 4885--4889, [\href{http://arxiv.org/abs/gr-qc/9710109}{{\sf
  arXiv:gr-qc/9710109}}],
  [\href{http://dx.doi.org/10.1103/PhysRevD.57.4885}{{\sf
  doi:10.1103/PhysRevD.57.4885}}].

\bibitem{Clement:1998nk}
G.~Clement, {\it {Selfgravitating cosmic rings}},  {\sl Phys.Lett.} {\bf B449}
  (1999) 12--16, [\href{http://arxiv.org/abs/gr-qc/9808082}{{\sf
  arXiv:gr-qc/9808082}}],
  [\href{http://dx.doi.org/10.1016/S0370-2693(99)00079-9}{{\sf
  doi:10.1016/S0370-2693(99)00079-9}}].

\bibitem{Bogush:2020lkp}
I.~Bogush and D.~Gal'tsov, {\it {Generation of rotating solutions in
  Einstein-scalar gravity}},  {\sl Phys. Rev. D} {\bf 102} (2020), no.~12
  124006, [\href{http://arxiv.org/abs/2001.02936}{{\sf arXiv:2001.02936}}],
  [\href{http://dx.doi.org/10.1103/PhysRevD.102.124006}{{\sf
  doi:10.1103/PhysRevD.102.124006}}].

\bibitem{Newman:1965tw}
E.~T. Newman and A.~I. Janis, {\it {Note on the Kerr spinning particle
  metric}},  {\sl J. Math. Phys.} {\bf 6} (1965) 915--917,
  [\href{http://dx.doi.org/10.1063/1.1704350}{{\sf doi:10.1063/1.1704350}}].

\bibitem{Zipoy}
D.~Zipoy, {\it {Topology of some spheroidal metrics}},  {\sl J.Math.Phys.} {\bf
  7} (1966) 1137--1143, [\href{http://dx.doi.org/10.1063/1.1705005}{{\sf
  doi:10.1063/1.1705005}}].

\bibitem{Voorhees:1971wh}
B.~H. Voorhees, {\it {Static axially symmetric gravitational fields}},  {\sl
  Phys.Rev.} {\bf D2} (1970) 2119--2122,
  [\href{http://dx.doi.org/10.1103/PhysRevD.2.2119}{{\sf
  doi:10.1103/PhysRevD.2.2119}}].

\bibitem{Gibbons:2017djb}
G.~W. Gibbons and M.~S. Volkov, {\it {Zero mass limit of Kerr spacetime is a
  wormhole}},  {\sl Phys. Rev. D} {\bf 96} (2017), no.~2 024053,
  [\href{http://arxiv.org/abs/1705.07787}{{\sf arXiv:1705.07787}}],
  [\href{http://dx.doi.org/10.1103/PhysRevD.96.024053}{{\sf
  doi:10.1103/PhysRevD.96.024053}}].

\bibitem{Eris}
A.~Eris and M.~Gurses, {\it {Stationary axially-symmetric solutions of
  Einstein-Maxwell massless scalar field equations}},  {\sl Journ.Math.Phys.}
  {\bf 18} (1977) 1303--1304, [\href{http://dx.doi.org/10.1063/1.523419}{{\sf
  doi:10.1063/1.523419}}].

\bibitem{Heusler1996}
M.~Heusler, {\em Black Hole Uniqueness Theorems}.
\newblock Cambridge University Press, 1996.

\bibitem{Ernst}
F.~J. Ernst, {\it {New formulation of the axially symmetric gravitational
  potential}},  {\sl Phys.Rev.} {\bf 167} (1968) 1175--1167,
  [\href{http://dx.doi.org/10.1103/PhysRev.167.1175}{{\sf
  doi:10.1103/PhysRev.167.1175}}].

\bibitem{Carter:1968rr}
B.~Carter, {\it {Global structure of the Kerr family of gravitational fields}},
   {\sl Phys.Rev.} {\bf 174} (1968) 1559--1571,
  [\href{http://dx.doi.org/10.1103/PhysRev.174.1559}{{\sf
  doi:10.1103/PhysRev.174.1559}}].

\bibitem{Tomimatsu:1972zz}
A.~Tomimatsu and H.~Sato, {\it {New exact solution for the gravitational field
  of a spinning mass}},  {\sl Phys. Rev. Lett.} {\bf 29} (1972) 1344--1345,
  [\href{http://dx.doi.org/10.1103/PhysRevLett.29.1344}{{\sf
  doi:10.1103/PhysRevLett.29.1344}}].

\bibitem{TS}
A.~Tomimatsu and H.~Sato, {\it {New series of exact solutions for gravitational
  fields of a spinning masses}},  {\sl Progr.Theor.Phys.} {\bf 50} (1973)
  95--110, [\href{http://dx.doi.org/10.1143/PTP.50.95}{{\sf
  doi:10.1143/PTP.50.95}}].

\bibitem{Cosgrove-I}
C.~M. Cosgrove, {\it {New family of exact stationary axisymmetric gravitational
  fields generalising the Tomimatsu-Sato solutions}},  {\sl Journ.Math.Phys.}
  {\bf 10} (1977) 1481--1524,
  [\href{http://dx.doi.org/10.1088/0305-4470/10/9/010}{{\sf
  doi:10.1088/0305-4470/10/9/010}}].

\bibitem{Cosgrove-II}
C.~M. Cosgrove, {\it {Limits of the generalised Tomimatsu-Sato gravitational
  field}},  {\sl Journ.Math.Phys.} {\bf 10} (1977) 2093--2015,
  [\href{http://dx.doi.org/10.1088/0305-4470/10/12/017}{{\sf
  doi:10.1088/0305-4470/10/12/017}}].

\bibitem{Cosgrove-III}
C.~M. Cosgrove, {\it {A new formulation of the field equations for the
  stationary axisymmetric vacuum gravitational field. I. General theory}},
  {\sl Journ.Math.Phys.} {\bf 11} (1978) 2389--2404,
  [\href{http://dx.doi.org/10.1088/0305-4470/11/12/007}{{\sf
  doi:10.1088/0305-4470/11/12/007}}].

\bibitem{Cosgrove-IV}
C.~M. Cosgrove, {\it {A new formulation of the field equations for the
  stationary axisymmetric vacuum gravitational field. II. Separable
  solutions}},  {\sl Journ.Math.Phys.} {\bf 11} (1978) 2405--2430,
  [\href{http://dx.doi.org/10.1088/0305-4470/11/12/008}{{\sf
  doi:10.1088/0305-4470/11/12/008}}].

\bibitem{Hori-0}
S.~Hori, {\it {On the exact solution of Tomimatsu-Sato family for an arbitrary
  integral value of the deformation parameter}},  {\sl Prog. Theor. Phys.} {\bf
  59} (1978) 1870, [\href{http://dx.doi.org/10.1143/PTP.59.1870}{{\sf
  doi:10.1143/PTP.59.1870}}]. [Erratum: Prog.Theor.Phys. 61, 365 (1979)].

\bibitem{Hori-I}
S.~Hori, {\it {Generalization of Tomimatsu-Sato solutions. I}},  {\sl Prog.
  Theor. Phys.} {\bf 95} (1996) 65--70,
  [\href{http://dx.doi.org/10.1143/PTP.95.65}{{\sf doi:10.1143/PTP.95.65}}].

\bibitem{Hori-II}
S.~Hori, {\it {Generalization of Tomimatsu-Sato solutions. II}},  {\sl Prog.
  Theor. Phys.} {\bf 95} (1996) 557--564,
  [\href{http://dx.doi.org/10.1143/PTP.95.557}{{\sf doi:10.1143/PTP.95.557}}].

\bibitem{Hori-III}
S.~Hori, {\it {Generalization of Tomimatsu-Sato solutions. III}},  {\sl Prog.
  Theor. Phys.} {\bf 95} (1996) 1097--1120,
  [\href{http://dx.doi.org/10.1143/PTP.95.1097}{{\sf
  doi:10.1143/PTP.95.1097}}].

\bibitem{Hori-IV}
S.~Hori, {\it {Generalization of Tomimatsu-Sato solutions. IV}},  {\sl Prog.
  Theor. Phys.} {\bf 96} (1996) 327--345,
  [\href{http://dx.doi.org/10.1143/PTP.96.327}{{\sf doi:10.1143/PTP.96.327}}].

\bibitem{Manko}
V.~Manko and C.~Moreno, {\it {Extension of the parameter space in the
  Tomimatsu-Sato solutions}},  {\sl Mod.Phys.Lett.} {\bf A 12} (1997) 613--617,
  [\href{http://dx.doi.org/10.1142/S0217732397000637}{{\sf
  doi:10.1142/S0217732397000637}}].

\bibitem{Clement:2015cxa}
G.~Cl\'ement, D.~Gal'tsov, and M.~Guenouche, {\it {Rehabilitating space-times
  with NUTs}},  {\sl Phys. Lett. B} {\bf 750} (2015) 591--594,
  [\href{http://arxiv.org/abs/1508.07622}{{\sf arXiv:1508.07622}}],
  [\href{http://dx.doi.org/10.1016/j.physletb.2015.09.074}{{\sf
  doi:10.1016/j.physletb.2015.09.074}}].

\bibitem{Clement:2015aka}
G.~Cl\'ement, D.~Gal'tsov, and M.~Guenouche, {\it {NUT wormholes}},  {\sl
  Phys.Rev.} {\bf D93} (2016), no.~2 024048,
  [\href{http://arxiv.org/abs/1509.07854}{{\sf arXiv:1509.07854}}],
  [\href{http://dx.doi.org/10.1103/PhysRevD.93.024048}{{\sf
  doi:10.1103/PhysRevD.93.024048}}]. [Phys. Rev.D93,024048(2016)].

\bibitem{Shinkai:2002gv}
H.-a. Shinkai and S.~A. Hayward, {\it {Fate of the first traversible wormhole:
  Black hole collapse or inflationary expansion}},  {\sl Phys. Rev. D} {\bf 66}
  (2002) 044005, [\href{http://arxiv.org/abs/gr-qc/0205041}{{\sf
  arXiv:gr-qc/0205041}}],
  [\href{http://dx.doi.org/10.1103/PhysRevD.66.044005}{{\sf
  doi:10.1103/PhysRevD.66.044005}}].

\bibitem{Belinski:2001ph}
V.~Belinski and E.~Verdaguer, {\em {Gravitational solitons}}.
\newblock Cambridge Monographs on Mathematical Physics. Cambridge University
  Press, 2005.

\end{thebibliography}

\providecommand{\href}[2]{#2}\begingroup\raggedright\endgroup

\end{document}